\pdfoutput=1
\documentclass{toc}%

\tocdetails{%
 volume = 1, number = 1, year=2005, firstpage=1, 
 received={June 29, 2004},
 published={February 9, 2005},
 doi={10.4086/toc.2005.v001a001}
}

\runningtitle{Limitations of Quantum Advice and One-way Communication}
\runningauthor{Scott Aaronson}
\copyrightauthor{Scott Aaronson}

\begin{document}
\begin{frontmatter}[classification=float]

\title{Limitations of Quantum Advice and One-Way Communication}
\tocpdftitle{Limitations of Quantum Advice and One-Way Communication}
\tocpdfauthor{Scott Aaronson}

\author[saaronson]{Scott Aaronson\thanks{Work supported by an
NSF Graduate Fellowship.}}

\runningauthor{Scott Aaronson}
\runningtitle{Limitations of Quantum Advice and One-Way Communication} %

\begin{abstract}
Although a quantum state requires exponentially many classical bits to
describe, the laws of quantum mechanics impose severe restrictions on
how that
state can be accessed. \ This paper shows in three settings that quantum
messages have only limited advantages over classical ones.

First, we show that $\mathsf{BQP/qpoly}\subseteq\mathsf{PP/poly}$, where
$\mathsf{BQP/qpoly}$\ is the class of problems solvable in quantum
polynomial
time, given a polynomial-size \textquotedblleft quantum advice
state\textquotedblright\ that depends only on the input length. \ This
resolves a question of Buhrman, and means that we should not hope for an
unrelativized separation between quantum and classical advice. \ Underlying
our complexity result is a general new relation between deterministic and
quantum one-way communication complexities, which applies to partial as well
as total functions.

Second, we construct an oracle relative to which $\mathsf{NP}\not
\subset \mathsf{BQP/qpoly}$. \ To do so, we use the polynomial
method to give the first correct proof of a \textit{direct product
theorem}\ for quantum search. \ This theorem has other applications;
for example, it can be used to fix a result of Klauck about quantum
time-space tradeoffs for sorting.

Third, we introduce a new \textit{trace distance method} for proving lower
bounds on quantum one-way communication complexity. \ Using this method, we
obtain optimal quantum lower bounds for two problems of Ambainis, for
which no
nontrivial lower bounds were previously known even for classical
randomized protocols.

A preliminary version of this paper appeared in the 2004 Conference
on Computational Complexity (CCC).
\end{abstract}

\tocams{81P68, 81P05, 68Q10, 68Q15, 42A05} \tocacm{F.1.2, F.1.3}
\tockeywords{quantum computation, advice, relativized complexity,
direct product theorem, nonuniform, BQP, PP, communication,
polynomial method, oracle}
\end{frontmatter}

\section{Introduction\label{INTRO}}

How many classical bits can \textquotedblleft really\textquotedblright\ be
encoded into $n$ qubits? \ Is it $n$, because of Holevo's Theorem
\cite{holevo}; $2n$, because of dense quantum coding \cite{bw}\ and quantum
teleportation \cite{bbcjpy}; exponentially many, because of quantum
fingerprinting \cite{bcww}; or infinitely many, because amplitudes are
continuous? \ The best general answer to this question is probably
\textit{mu}, the Zen word that \textquotedblleft unasks\textquotedblright\ a
question.\footnote{Another \textit{mu}-worthy question is, \textquotedblleft
Where does the power of quantum computing come from? \ Superposition?
\ Interference? \ The large size of Hilbert space?\textquotedblright}

To a computer scientist, however, it is natural to formalize the question in
terms of \textit{quantum one-way communication complexity}
\cite{bjk,bcww,klauck:cc,yao:fing}. \ The setting is as follows: Alice
has an
$n$-bit string $x$, Bob has an $m$-bit string $y$, and together they wish to
evaluate $f\left(  x,y\right)  $\ where $f:\left\{  0,1\right\}  ^{n}%
\times\left\{  0,1\right\}  ^{m}\rightarrow\left\{  0,1\right\}  $\ is a
Boolean function. \ After examining her input $x=x_{1}\ldots x_{n}$,
Alice can
send a single quantum message $\rho_{x}$\ to Bob, whereupon Bob, after
examining his input $y=y_{1}\ldots y_{m}$, can choose some basis in which to
measure $\rho_{x}$. \ He must then output a claimed value for $f\left(
x,y\right)  $. \ We are interested in how long Alice's message needs to be,
for Bob to succeed with high probability on any $x,y$ pair. \ Ideally the
length will be much smaller than if Alice had to send a classical message.

Communication complexity questions have been intensively studied in
theoretical computer science (see the book of Kushilevitz and Nisan
\cite{kn}\ for example). \ In both the classical and quantum cases, though,
most attention has focused on \textit{two-way }communication, meaning that
Alice and Bob get to send messages back and forth. \ We believe that the
study
of one-way quantum communication presents two main advantages. \ First, many
open problems about two-way communication look gruesomely difficult---for
example, are the randomized and quantum communication complexities of every
total Boolean function polynomially related? \ We might gain insight into
these problems by tackling their one-way analogues first. \ And second,
because of its greater simplicity, the one-way model more directly addresses
our opening question: how much \textquotedblleft useful
stuff\textquotedblright\ can be packed into a quantum state? \ Thus, results
on one-way communication fall into the quantum information theory tradition
initiated by Holevo \cite{holevo}\ and others, as much as the communication
complexity tradition initiated by Yao \cite{yao:cc}.

Related to quantum one-way communication is the notion of \textit{quantum
advice}.\ \ As pointed out by Nielsen and Chuang \cite[p.203]{nc}, there
is no
compelling physical reason to assume that the starting state of a quantum
computer is a computational basis state:\footnote{One might object that the
starting state is itself the outcome of some computational process, which
began no earlier than the Big Bang. \ However, (1) for all we know highly
entangled states were created in the Big Bang, and (2) $14$ billion
years is a
long time.}

\begin{quote}
[W]e know that many systems in Nature `prefer' to sit in highly entangled
states of many systems; might it be possible to exploit this preference to
obtain extra computational power?\ \ It might be that having access to
certain
states allows particular computations to be done much more easily than if we
are constrained to start in the computational basis.
\end{quote}

One way to interpret Nielsen and Chuang's provocative question is as
follows.
\ Suppose we could request the \textit{best possible} starting state for a
quantum computer, knowing the language to be decided and the input
length $n$
but not knowing the input itself.\footnote{If we knew the input, we would
simply request a starting state that contains the right answer!} \
Denote the
class of languages that we could then decide by
$\mathsf{BQP/qpoly}$---meaning
quantum polynomial time, given an arbitrarily-entangled but polynomial-size
quantum advice state.\footnote{$\mathsf{BQP/qpoly}$ might remind readers
of a
better-studied class called $\mathsf{QMA}$ (Quantum Merlin-Arthur). \ But
there are two key differences: first, advice can be trusted while proofs
cannot; second, proofs can be tailored to a particular input while advice
cannot.} \ How powerful is this class? \ If $\mathsf{BQP/qpoly}$ contained
(for example) the $\mathsf{NP}$-complete problems, then we would need to
rethink our most basic assumptions about the power of quantum computing.
\ We
will see later that quantum advice is closely related to quantum one-way
communication, since we can think of an advice state as a one-way
message sent
to an algorithm by a benevolent \textquotedblleft advisor.\textquotedblright

This paper is about the \textit{limitations} of quantum advice and one-way
communication. \ It presents three contributions which are basically
independent of one another.

First, %
\secref{1WAYSIM} shows that $D^{1}\left(  f\right)  =O\left(
mQ_{2}^{1}\left(  f\right)  \log Q_{2}^{1}\left(  f\right)  \right)  $\ for
any Boolean function $f$, partial or total. \ Here $D^{1}\left(  f\right)
$\ is deterministic one-way communication complexity, $Q_{2}^{1}\left(
f\right)  $\ is bounded-error one-way quantum communication complexity, and
$m$ is the length of Bob's input. \ Intuitively, whenever the set of Bob's
possible inputs is not too large, Alice can send him a short classical
message
that lets him learn the outcome of any measurement he would have wanted to
make on the quantum message $\rho_{x}$. \ It is interesting that a slightly
tighter bound for total functions---$D^{1}\left(  f\right)  =O\left(
mQ_{2}^{1}\left(  f\right)  \right)  $---follows easily from a result of
Klauck \cite{klauck:cc} together with a lemma of Sauer \cite{sauer} about
VC-dimension. \ However, the proof of the latter bound is highly
nonconstructive, and seems to fail for partial $f$.\nolinebreak

Using our communication complexity result, in %
\secref{ADVICESIM} we show
that $\mathsf{BQP/qpoly}\subseteq\mathsf{PP/poly}$---in other words,
$\mathsf{BQP}$\ with polynomial-size quantum advice can be simulated in
$\mathsf{PP}$\ with polynomial-size classical advice.\footnote{Here
$\mathsf{PP}$\ is Probabilistic Polynomial-Time, or the class of
languages for
which there exists a polynomial-time\ classical randomized algorithm that
accepts with probability greater than $1/2$\ if and only if an input $x$\ is
in the language. \ Also, given a complexity class $\mathsf{C}$, the class
$\mathsf{C/poly}$ consists of all languages decidable by a $\mathsf{C}%
$\ machine, given a polynomial-size classical advice string\ that
depends only
on the input length. \ See www.complexityzoo.com for more information about
standard complexity classes mentioned in this paper.} \ This resolves a
question of Harry Buhrman (personal communication), who asked whether
quantum
advice\ can be simulated in \textit{any} classical complexity class with
short
classical advice. \ A corollary of our containment is that we cannot hope to
show an unrelativized separation between quantum and classical advice (that
is, that $\mathsf{BQP/poly}\neq\mathsf{BQP/qpoly}$), without also
showing that
$\mathsf{PP}$\ does not have polynomial-size circuits.

What makes this result surprising is that, in the minds of many computer
scientists, a quantum state is basically an exponentially long vector.
\ Indeed, this belief seems to fuel skepticism of quantum computing (see
Goldreich \cite{goldreich:qc} for example). \ But given an exponentially
long
advice string, even a classical computer could decide any language
whatsoever.
\ So one might\ imagine\ na\"{\i}vely\ that quantum advice would let us
solve
problems that are not even recursively enumerable given classical advice
of a
similar size! \ The failure of this na\"{\i}ve intuition supports the view
that a quantum superposition over $n$-bit strings is \textquotedblleft more
similar\textquotedblright\ to a probability distribution over $n$-bit
strings
than to a $2^{n}$-bit string.

\begin{sloppypar}
Our second contribution, in %
\secref{ORACLE}, is an oracle relative to which $\mathsf{NP}$\ is
not contained in\ $\mathsf{BQP/qpoly}$. \ Underlying this oracle
separation is the first correct proof of a \textit{direct product
theorem} for quantum search.\ \ Given an $N$-item database with $K$
marked items, the direct product theorem says that if a quantum
algorithm makes $o\left(  \sqrt{N}\right)  $ queries, then the
probability that the algorithm finds all $K$ of the marked items
decreases exponentially in $K$. \ Notice that such a result does not
follow from any existing quantum lower bound. \ Earlier Klauck
\cite{klauck:ts}\ claimed a weaker direct product theorem, based on
the hybrid method of Bennett et al. \cite{bbbv}, in a paper on
quantum time-space tradeoffs for sorting. \ Unfortunately, Klauck's
proof contained a bug. \ Our proof uses the polynomial method of
Beals et al. \cite{bbcmw},\ with the novel twist that we examine all
\textit{higher} derivatives of a polynomial (not just the first
derivative). \ Our proof has already been improved by Klauck,
\v{S}palek, and de Wolf \cite{ksw}, who were able to recover and
even extend Klauck's original results about quantum sorting.
\end{sloppypar}

Our final contribution, in %
\secref{TDIST}, is a new \textit{trace
distance method} for proving lower bounds on quantum one-way communication
complexity. \ Previously there was only one basic lower bound technique: the
VC-dimension method of Klauck \cite{klauck:cc}, which relied on lower bounds
for quantum random access codes due to Ambainis et al. \cite{antv}\ and
Nayak
\cite{nayak}. \ Using VC-dimension one can show, for example, that $Q_{2}%
^{1}\left(  \operatorname*{DISJ}\right)  =\Omega\left(  n\right)  $,
where the
\textit{disjointness function}\ $\operatorname*{DISJ}:\left\{  0,1\right\}
^{n}\times\left\{  0,1\right\}  ^{n}\rightarrow\left\{  0,1\right\}  $\ is
defined by \ $\operatorname*{DISJ}\left(  x,y\right)  =1$ if and only if
$x_{i}y_{i}=0$\ for all $i\in\left\{  1,\ldots,n\right\}  $.

For some problems, however, the VC-dimension method yields no nontrivial
quantum lower bound. \ Seeking to make this point vividly, Ambainis
posed the
following problem. \ Alice is given two elements $x,y$\ of a finite field
$\mathbb{F}_{p}$ (where $p$ is prime); Bob is given another two elements
$a,b\in\mathbb{F}_{p}$. \ Bob's goal is to output $1$ if $y\equiv ax+b\left(
\operatorname{mod}p\right)  $ and $0$ otherwise. \ For this problem, the
VC-dimension method yields no randomized \textit{or} quantum lower bound
better than constant. \ On the other hand, the well-known fingerprinting
protocol for the equality function \cite{ry}\ seems to fail for Ambainis'
problem, because of the interplay between addition and multiplication. \
So it
is natural to conjecture that the randomized and even quantum one-way
complexities are $\Theta\left(  \log p\right)  $---that is, that no
nontrivial
protocol exists for this problem.

Ambainis posed a second problem in the same spirit. \ Here Alice is given
$x\in\left\{  1,\ldots,N\right\}  $, Bob is given $y\in\left\{  1,\ldots
,N\right\}  $, and both players know a subset $S\subset\left\{  1,\ldots
,N\right\}  $. \ Bob's goal is to decide whether $x-y\in S$\ where
subtraction
is modulo $N$. \ The conjecture is that if $S$ is chosen uniformly at random
with $\left\vert S\right\vert $ about $\sqrt{N}$, then with high probability
the randomized and quantum one-way complexities are both $\Theta\left(  \log
N\right)  $.

Using our trace distance method, we are able to show optimal quantum lower
bounds for both of Ambainis' problems. \ Previously, no nontrivial lower
bounds were known even for randomized protocols. \ The key idea is to
consider
two probability distributions over Alice's quantum message $\rho_{x}$. \ The
first distribution corresponds to $x$ chosen uniformly at random; the second
corresponds to $x$ chosen uniformly conditioned on $f\left(  x,y\right)
=1$.
\ These distributions give rise to two mixed states $\rho$\ and $\rho_{y}$,
which Bob must be able to distinguish with non-negligible bias assuming
he can
evaluate $f\left(  x,y\right)  $. \ We then show an upper bound on the trace
distance $\left\Vert \rho-\rho_{y}\right\Vert _{\operatorname*{tr}}$, which
implies that Bob cannot distinguish the distributions.

\autoref{vardist}\ gives a very general condition under which our trace
distance method works;\ Corollaries \ref{coset}\ and \ref{subset}
then show
that the condition is satisfied for Ambainis' two problems. \ Besides
showing
a significant limitation of the VC-dimension method, we hope our new
method is
a non-negligible step towards proving that $R_{2}^{1}\left(  f\right)
=O\left(  Q_{2}^{1}\left(  f\right)  \right)  $\ for all total Boolean
functions $f$, where $R_{2}^{1}\left(  f\right)  $\ is randomized
one-way complexity. \ We conclude in %
\secref{OPEN}\ with some open problems.

This paper is a moderately revised version of an extended abstract
that appeared in CCC 2004 \cite{aar:advconf}. \ The proofs of
Theorems \ref{partialthm}, \ref{qpoly}, \ref{vardist}, \ref{coset},
and \ref{randset} have been written out in more detail; and
discussions have been added to Sections \ref{ADVICE} and
\ref{ADVICESIM}, about the group membership problem and the class
$\mathsf{PQP/qpoly}$ respectively. \ Also, an error has been fixed
in Section \ref{ORACLE}: the direct product theorem in
\cite{aar:advconf} based on Bernstein's inequality is incorrect. \
Fortunately, the easier version based on V. A. Markov's inequality
is still perfectly sufficient to show $\mathsf{NP}\not \subset
\mathsf{BQP/qpoly}$ relative to an oracle; and in any case, the
Bernstein's version has been superseded by the results of Klauck,
\v{S}palek, and de Wolf \cite{ksw}.

\section{Preliminaries\label{PRELIM}}

This section reviews basic definitions and results about quantum one-way
communication (in %
\secref{1WAY}) and quantum advice (in %
\secref{ADVICE}); then %
\secref{GOODASNEW}\ proves a quantum information
lemma that will be used throughout the paper.

\subsection{Quantum One-Way Communication\label{1WAY}}

Following standard conventions, we denote by $D^{1}\left(  f\right)  $\ the
deterministic one-way complexity of $f$, or the minimum number of bits that
Alice must send if her message is a function of $x$. \ Also,
$R_{2}^{1}\left(
f\right)  $, the bounded-error randomized one-way complexity, is the minimum
$k$ such that for every $x,y$, if Alice sends Bob a $k$-bit message
drawn from
some distribution $\mathcal{D}_{x}$, then Bob can output a bit $a$ such that
$a=f\left(  x,y\right)  $\ with probability at least $2/3$. \ (The subscript
$2$ means that the error is two-sided.) \ The zero-error randomized
complexity
$R_{0}^{1}\left(  f\right)  $\ is similar, except that Bob's answer can
never
be wrong: he must output $f\left(  x,y\right)  $\ with probability at least
$1/2$ and otherwise declare failure.

The bounded-error quantum one-way complexity $Q_{2}^{1}\left(  f\right)
$\ is
the minimum $k$ such that, if Alice sends Bob a mixed state $\rho_{x}$\
of $k$
qubits, there exists a joint measurement of $\rho_{x}$\ and $y$ enabling Bob
to output an $a$ such that $a=f\left(  x,y\right)  $\ with probability at
least $2/3$. \ The zero-error and exact complexities $Q_{0}^{1}\left(
f\right)  $\ and $Q_{E}^{1}\left(  f\right)  $\ are defined analogously.
\ Requiring Alice's message to be a pure state would increase these
complexities by at most a factor of $2$, since by Kraus' Theorem, every
$k$-qubit mixed state can be realized as half of a $2k$-qubit pure state.
\ (Winter \cite{winter}\ has shown that this factor of $2$ is tight.) \ See
Klauck \cite{klauck:cc}\ for more detailed definitions of quantum and
classical one-way communication complexity measures.

It is immediate that $D^{1}\left(  f\right)  \geq R_{0}^{1}\left(  f\right)
\geq R_{2}^{1}\left(  f\right)  \geq Q_{2}^{1}\left(  f\right)  $, that
$R_{0}^{1}\left(  f\right)  \geq Q_{0}^{1}\left(  f\right)  \geq Q_{2}%
^{1}\left(  f\right)  $, and that $D^{1}\left(  f\right)  \geq Q_{E}%
^{1}\left(  f\right)  $. \ Also, for total $f$, Duri\v{s} et al. \cite{dhrs}
showed that $R_{0}^{1}\left(  f\right)  =\Theta\left(  D^{1}\left(  f\right)
\right)  $, while Klauck \cite{klauck:cc}\ showed that $Q_{E}^{1}\left(
f\right)  =D^{1}\left(  f\right)  $\ and that $Q_{0}^{1}\left(  f\right)
=\Theta\left(  D^{1}\left(  f\right)  \right)  $. \ In other words,
randomized
and quantum messages yield no improvement for total functions if we are
unwilling to tolerate a bounded probability of error. \ This remains\ true
even if Alice and Bob share arbitrarily many EPR pairs \cite{klauck:cc}.
\ As
is often the case, the situation is dramatically different for partial
functions: there it is easy to see that $R_{0}^{1}\left(  f\right)  $\
can be
constant\ even though $D^{1}\left(  f\right)  =\Omega\left(  n\right)
$: let
$f\left(  x,y\right)  =1$\ if $x_{1}y_{1}+\cdots+x_{n/2}y_{n/2}\geq
n/4$\ and\ $x_{n/2+1}y_{n/2+1}+\cdots+x_{n}y_{n}=0$ and $f\left(  x,y\right)
=0$\ if $x_{1}y_{1}+\cdots+x_{n/2}y_{n/2}=0$\ and\ $x_{n/2+1}y_{n/2+1}%
+\cdots+x_{n}y_{n}\geq n/4$, promised that one of these is the case.

Moreover, Bar-Yossef, Jayram, and Kerenidis \cite{bjk}\ have \textit{almost}
shown that $Q_{E}^{1}\left(  f\right)  $\ can be exponentially smaller than
$R_{2}^{1}\left(  f\right)  $. \ In particular, they proved that separation
for a \textit{relation}, meaning a problem for which Bob has many possible
valid outputs. \ For a partial function $f$ based on their relation,
they also
showed that $Q_{E}^{1}\left(  f\right)  =\Theta\left(  \log n\right)
$\ whereas $R_{0}^{1}\left(  f\right)  =\Theta\left(  \sqrt{n}\right)
$; and
they conjectured (but did not prove) that $R_{2}^{1}\left(  f\right)
=\Theta\left(  \sqrt{n}\right)  $.

\subsection{Quantum Advice\label{ADVICE}}

Informally, $\mathsf{BQP/qpoly}$ is the class of languages decidable in
polynomial time on a quantum computer, given a polynomial-size quantum
advice
state that depends only on the input length. \ We now make the
definition more formal.

\begin{definition}
\label{qpolydef}A language $L$ is in $\mathsf{BQP/qpoly}$\ if there exists a
polynomial-size quantum circuit family $\left\{  C_{n}\right\}  _{n\geq1}$,
and a polynomial-size family of quantum states $\left\{  \left|  \psi
_{n}\right\rangle \right\}  _{n\geq1}$, such that for all $x\in\left\{
0,1\right\}  ^{n}$,

\begin{enumerate}
\item[(i)] If $x\in L\ $then $q\left(  x\right)  \geq2/3$, where $q\left(
x\right)  $\ is\ the probability that the first qubit is measured to be
$\left|  1\right\rangle $, after $C_{n}$\ is applied to the starting state
$\left|  x\right\rangle \otimes\left|  0\cdots0\right\rangle \otimes\left|
\psi_{n}\right\rangle $.

\item[(ii)] If $x\notin L$\ then$\ q\left(  x\right)  \leq1/3$.\footnote{If
the starting state is $\left\vert x\right\rangle \otimes\left\vert
0\cdots0\right\rangle \otimes\left\vert \varphi\right\rangle $\ for some
$\left\vert \varphi\right\rangle \neq\left\vert \psi_{n}\right\rangle $,
then
we do not require the acceptance probability to lie in $\left[  0,1/3\right]
\cup\left[  2/3,1\right]  $. \ Therefore, what we call $\mathsf{BQP/qpoly}%
$\ corresponds to what Nishimura and Yamakami \cite{ny}\ call
$\mathsf{BQP/}%
^{\mathsf{\ast}}\mathsf{Qpoly}$. \ Also, it does not matter whether the
circuit family $\left\{  C_{n}\right\}  _{n\geq1}$\ is uniform, since we are
giving it advice anyway.}
\end{enumerate}
\end{definition}

\begin{sloppypar}
The central open question about $\mathsf{BQP/qpoly}$\ is whether it equals
$\mathsf{BQP/poly}$, or $\mathsf{BQP}$ with polynomial-size
\textit{classical}
advice. \ We do have a candidate for an oracle problem separating the two
classes: the \textit{group membership problem} of Watrous \cite{watrous},
which we describe for completeness. \ Let $G_{n}$\ be a black box
group\footnote{In other words, we have a quantum oracle available that given
$x,y\in G_{n}$\ outputs $xy$ (i.e. exclusive-OR's $xy$ into an answer
register), and that given $x\in G_{n}$\ outputs $x^{-1}$.} whose
elements are
uniquely labeled by $n$-bit strings,\ and let $H_{n}$\ be a subgroup of
$G_{n}$. \ Both $G_{n}$\ and $H_{n}$\ depend only on the input length\
$n$, so
we can assume that a nonuniform algorithm knows generating sets for both of
them. Given an element $x\in G_{n}$ as input, the problem is to decide
whether $x\in H_{n}$.
\end{sloppypar}

If $G_{n}$\ is \textquotedblleft sufficiently nonabelian\textquotedblright%
\ and $H_{n}$\ is exponentially large, we do not know how to solve this
problem in $\mathsf{BQP}$\ or even $\mathsf{BQP/poly}$. \ On the other hand,
we can solve it in $\mathsf{BQP/qpoly}$\ as follows. \ Let our quantum
advice
state be an equal superposition over all elements of $H_{n}$:%
\[
\left\vert H_{n}\right\rangle =\frac{1}{\sqrt{\left\vert H_{n}\right\vert }%
}\sum_{y\in H_{n}}\left\vert y\right\rangle
\]
We can transform $\left\vert H_{n}\right\rangle $\ into%
\[
\left\vert xH_{n}\right\rangle =\frac{1}{\sqrt{\left\vert
H_{n}\right\vert }%
}\sum_{y\in H_{n}}\left\vert xy\right\rangle
\]
by mapping $\left\vert y\right\rangle \left\vert 0\right\rangle $ to
$\left\vert y\right\rangle \left\vert xy\right\rangle $\ to $\left\vert
y\oplus x^{-1}xy\right\rangle \left\vert xy\right\rangle =\left\vert
0\right\rangle \left\vert xy\right\rangle $\ for each $y\in H_{n}$. Our
algorithm will first prepare the state $\left(  \left\vert 0\right\rangle
\left\vert H_{n}\right\rangle +\left\vert 1\right\rangle \left\vert
xH_{n}\right\rangle \right)  /\sqrt{2}$,\ then apply a Hadamard gate to the
first qubit, and finally measure the first qubit in the standard basis, in
order to distinguish the cases $\left\vert H_{n}\right\rangle =\left\vert
xH_{n}\right\rangle $\ and $\left\langle H_{n}|xH_{n}\right\rangle =0$\ with
constant bias. \ The first case occurs whenever $x\in H_{n}$, and the second
occurs whenever $x\notin H_{n}$.

Although the group membership problem provides intriguing evidence for the
power of quantum advice, we have no idea how to show that it is not also
solvable using classical advice. \ Indeed, apart from a result of Nishimura
and Yamakami \cite{ny}\ that $\mathsf{EESPACE}\not \subset
\mathsf{BQP/qpoly}%
$, essentially nothing was known about the class $\mathsf{BQP/qpoly}$\
before
the present work.

\subsection{The Almost As Good As New Lemma\label{GOODASNEW}}

The following simple lemma, which was implicit in \cite{antv}, is used three
times in this paper---in Theorems \ref{partialthm}, \ref{qpoly}, and
\ref{npqpoly}. \ It says that, if the outcome of measuring a quantum state
$\rho$\ could be predicted with near-certainty given knowledge of
$\rho$, then
measuring $\rho$ will damage it only slightly. \ Recall that the trace
distance $\left\|  \rho-\sigma\right\|  _{\operatorname*{tr}}$\ between two
mixed states $\rho$\ and $\sigma$ equals $\frac{1}{2}\sum_{i}\left|
\lambda_{i}\right|  $, where $\lambda_{1},\ldots,\lambda_{N}$\ are the
eigenvalues of $\rho-\sigma$.

\begin{lemma}
\label{tracedist}Suppose a $2$-outcome measurement of a mixed state $\rho$
yields outcome $0$ with probability $1-\varepsilon$. \ Then after the
measurement, we can recover a state $\widetilde{\rho}$\ such that $\left\|
\widetilde{\rho}-\rho\right\|  _{\operatorname*{tr}}\leq\sqrt{\varepsilon}$.
\ This is true even if the measurement is a POVM (that is, involves
arbitrarily many ancilla qubits).
\end{lemma}

\begin{proof}
Let $\left|  \psi\right\rangle $\ be a purification of the entire system
($\rho$ plus ancilla). \ We can represent any measurement as a unitary $U$
applied to $\left|  \psi\right\rangle $, followed by a $1$-qubit
measurement.
\ Let $\left|  \varphi_{0}\right\rangle $\ and $\left|  \varphi_{1}%
\right\rangle $\ be the two possible pure states after the measurement; then
$\left\langle \varphi_{0}|\varphi_{1}\right\rangle =0$\ and $U\left|
\psi\right\rangle =\alpha\left|  \varphi_{0}\right\rangle +\beta\left|
\varphi_{1}\right\rangle $ for some $\alpha,\beta$\ such that $\left|
\alpha\right|  ^{2}=1-\varepsilon$\ and $\left|  \beta\right|  ^{2}%
=\varepsilon$. \ Writing the measurement result as $\sigma=\left(
1-\varepsilon\right)  \left|  \varphi_{0}\right\rangle \left\langle
\varphi_{0}\right|  +\varepsilon\left|  \varphi_{1}\right\rangle
\left\langle
\varphi_{1}\right|  $, it is easy to show that%
\[
\left\|  \sigma-U\left|  \psi\right\rangle \left\langle \psi\right|
U^{-1}\right\|  _{\operatorname*{tr}}=\sqrt{\varepsilon\left(  1-\varepsilon
\right)  }.
\]
So applying $U^{-1}$\ to $\sigma$,%
\[
\left\|  U^{-1}\sigma U-\left|  \psi\right\rangle \left\langle \psi\right|
\right\|  _{\operatorname*{tr}}=\sqrt{\varepsilon\left(
1-\varepsilon\right)
}.
\]
Let $\widetilde{\rho}$\ be the restriction of $U^{-1}\sigma U$\ to the
original qubits of $\rho$. \ Theorem 9.2 of Nielsen and Chuang \cite{nc}
shows
that tracing out a subsystem never increases trace distance, so $\left\|
\widetilde{\rho}-\rho\right\|  _{\operatorname*{tr}}\leq\sqrt{\varepsilon
\left(  1-\varepsilon\right)  }\leq\sqrt{\varepsilon}$.
\end{proof}

\section{Simulating Quantum Messages\label{1WAYSIM}}

Let $f:\left\{  0,1\right\}  ^{n}\times\left\{  0,1\right\}  ^{m}%
\rightarrow\left\{  0,1\right\}  $\ be a Boolean function. \ In this section
we first combine existing results to obtain the relation $D^{1}\left(
f\right)  =O\left(  mQ_{2}^{1}\left(  f\right)  \right)  $ for total
$f$, and
then prove using a new method that $D^{1}\left(  f\right)  =O\left(
mQ_{2}^{1}\left(  f\right)  \log Q_{2}^{1}\left(  f\right)  \right)  $\ for
all $f$ (partial or total).

Define the \textit{communication matrix} $M_{f}$\ to be a $2^{n}\times2^{m}%
$\ matrix with $f\left(  x,y\right)  $\ in the $x^{th}$\ row and $y^{th}%
$\ column.\ \ Then letting $\operatorname*{rows}\left(  f\right)  $\ be the
number of distinct rows in $M_{f}$, the following is immediate.

\begin{proposition}
\label{nrows}For total $f$,%
\begin{align*}
D^{1}\left(  f\right)   &  =\left\lceil \log_{2}\operatorname*{rows}\left(
f\right)  \right\rceil ,\\
Q_{2}^{1}\left(  f\right)   &  =\Omega\left(  \log\log\operatorname*{rows}%
\left(  f\right)  \right)  .
\end{align*}

\end{proposition}

Also, let the VC-dimension $\operatorname*{VC}\left(  f\right)  $\ equal the
maximum $k$ for which there exists a\ $2^{n}\times k$\ submatrix $M_{g}$\ of
$M_{f}$\ with $\operatorname*{rows}\left(  g\right)  =2^{k}$. \ Then Klauck
\cite{klauck:cc}\ observed the following, based on a lower bound for quantum
random access codes due to Nayak \cite{nayak}.

\begin{proposition}
[Klauck]\label{vc}$Q_{2}^{1}\left(  f\right)  =\Omega\left(
\operatorname*{VC}\left(  f\right)  \right)  $ for total $f$.
\end{proposition}

Now let $\operatorname*{cols}\left(  f\right)  $\ be the number of distinct
columns in $M_{f}$. \ Then %
\autoref{vc}\ yields the following general lower bound:

\begin{corollary}
\label{vccor}$D^{1}\left(  f\right)  =O\left(  mQ_{2}^{1}\left(  f\right)
\right)  $ for total $f$, where $m$ is the size of Bob's input.
\end{corollary}

\begin{proof}
It follows from a lemma of Sauer \cite{sauer}\ that%
\[
\operatorname*{rows}\left(  f\right)  \leq\sum_{i=0}^{\operatorname*{VC}%
\left(  f\right)  }\dbinom{\operatorname*{cols}\left(  f\right)  }{i}%
\leq\operatorname*{cols}\left(  f\right)  ^{\operatorname*{VC}\left(
f\right)  +1}.
\]
Hence $\operatorname*{VC}\left(  f\right)  \geq\log_{\operatorname*{cols}%
\left(  f\right)  }\operatorname*{rows}\left(  f\right)  -1$, so%
\begin{align*}
Q_{2}^{1}\left(  f\right)  =\Omega\left(  \operatorname*{VC}\left(  f\right)
\right)   &  =\Omega\left(  \frac{\log\operatorname*{rows}\left(  f\right)
}{\log\operatorname*{cols}\left(  f\right)  }\right) \\
&  =\Omega\left(  \frac{D^{1}\left(  f\right)  }{m}\right)  .
\end{align*}

\end{proof}

In particular, $D^{1}\left(  f\right)  $\ and $Q_{2}^{1}\left(  f\right)
$\ are polynomially related for total $f$, whenever Bob's input is
polynomially smaller than Alice's, and Alice's input is not
\textquotedblleft
padded.\textquotedblright\ More formally, $D^{1}\left(  f\right)  =O\left(
Q_{2}^{1}(  f)  ^{1/\left(  1-c\right)  }\right)  $ whenever
$m=O\left(  n^{c}\right)  $\ for some $c<1$ and $\operatorname*{rows}\left(
f\right)  =2^{n}$ (i.e. all rows of $M_{f}$\ are distinct). \ For then
$D^{1}\left(  f\right)  =n$ by %
\autoref{nrows}, and $Q_{2}^{1}\left(
f\right)  =\Omega\left(  D^{1}\left(  f\right)  /n^{c}\right)  =\Omega\left(
n^{1-c}\right)  $\ by %
\corref{vccor}.

\begin{sloppypar}
We now give a new method for replacing quantum messages by classical
ones when
Bob's input is small. \ Although the best bound we know how to obtain with
this method---$D^{1}\left(  f\right)  =O\left(  mQ_{2}^{1}\left(  f\right)
\log Q_{2}^{1}\left(  f\right)  \right)  $---is slightly weaker than the
$D^{1}\left(  f\right)  =O\left(  mQ_{2}^{1}\left(  f\right)  \right)  $ of
\corref{vccor}, our method works for \textit{partial} Boolean functions
as well as total ones.\ \ It also yields a (relatively) efficient
procedure by
which Bob can reconstruct Alice's quantum message, a fact we will exploit in
\secref{ADVICESIM}\ to show $\mathsf{BQP/qpoly}\subseteq\mathsf{PP/poly}%
$. \ By contrast, the method based on Sauer's Lemma seems to be
nonconstructive.
\end{sloppypar}

\begin{theorem}
\label{partialthm}$D^{1}\left(  f\right)  =O\left(  mQ_{2}^{1}\left(
f\right)  \log Q_{2}^{1}\left(  f\right)  \right)  $ for all $f$
(partial or total).
\end{theorem}

\begin{proof}
Let $f:\mathcal{D}\rightarrow\left\{  0,1\right\}  $\ be a partial Boolean
function with $\mathcal{D}\subseteq\left\{  0,1\right\}  ^{n}\times\left\{
0,1\right\}  ^{m}$, and for all $x\in\left\{  0,1\right\}  ^{n}$, let
$\mathcal{D}_{x}=\left\{  y\in\left\{  0,1\right\}  ^{m}:\left(  x,y\right)
\in\mathcal{D}\right\}  $. \ Suppose Alice can send Bob a quantum state with
$Q_{2}^{1}\left(  f\right)  $\ qubits, that enables him to compute $f\left(
x,y\right)  $\ for any $y\in\mathcal{D}_{x}$\ with error probability at most
$1/3$. \ Then she can also send him a boosted state $\rho$ with $K=O\left(
Q_{2}^{1}\left(  f\right)  \log Q_{2}^{1}\left(  f\right)  \right)  $
qubits,
such that for all $y\in\mathcal{D}_{x}$,%
\[
\left\vert P_{y}\left(  \rho\right)  -f\left(  x,y\right)  \right\vert
\leq\frac{1}{Q_{2}^{1}\left(  f\right)  ^{10}},
\]
where $P_{y}\left(  \rho\right)  $\ is the probability that some measurement
$\Lambda\left[  y\right]  $\ yields a `$1$'\ outcome when applied to $\rho$.
\ We can assume for simplicity that $\rho$\ is a pure state $\left\vert
\psi\right\rangle \left\langle \psi\right\vert $; as discussed in %
\secref{1WAY}, this increases the message length by at most a factor of $2$.

Let $\mathcal{Y}$\ be any subset of $\mathcal{D}_{x}$\ satisfying
$\left\vert
\mathcal{Y}\right\vert \leq Q_{2}^{1}\left(  f\right)  ^{2}$. \ Then
starting
with $\rho$, Bob can measure $\Lambda\left[  y\right]  $\ for each
$y\in\mathcal{Y}$ in lexicographic order, reusing the same message
state\ again and again but uncomputing whatever garbage he generates while
measuring. \ Let $\rho_{t}$\ be the state after the $t^{th}$\ measurement;
thus $\rho_{0}=\rho=\left\vert \psi\right\rangle \left\langle
\psi\right\vert
$. \ Since the probability that Bob outputs the wrong value of $f\left(
x,y\right)  $ on any given $y$ is at most $1/Q_{2}^{1}\left(  f\right)
^{10}%
$,\ %
\lemref{tracedist} implies that%
\[
\left\Vert \rho_{t}-\rho_{t-1}\right\Vert _{\operatorname*{tr}}\leq\sqrt
{\frac{1}{Q_{2}^{1}\left(  f\right)  ^{10}}}=\frac{1}{Q_{2}^{1}\left(
f\right)  ^{5}}.
\]
Since trace distance satisfies the triangle inequality, this in turn implies
that%
\[
\left\Vert \rho_{t}-\rho\right\Vert _{\operatorname*{tr}}\leq\frac{t}%
{Q_{2}^{1}\left(  f\right)  ^{5}}\leq\frac{1}{Q_{2}^{1}\left(  f\right)
^{3}%
}.
\]
Now imagine an \textquotedblleft ideal scenario\textquotedblright\ in which
$\rho_{t}=\rho$ for every $t$; that is, the measurements do not damage $\rho
$\ at all. \ Then the maximum bias with which Bob could distinguish the
actual
from the ideal scenario is%
\[
\left\Vert \rho_{0}-\rho\right\Vert _{\operatorname*{tr}}+\cdots+\left\Vert
\rho_{\left\vert \mathcal{Y}\right\vert -1}-\rho\right\Vert
_{\operatorname*{tr}}\leq\frac{\left\vert \mathcal{Y}\right\vert }{Q_{2}%
^{1}\left(  f\right)  ^{3}}\leq\frac{1}{Q_{2}^{1}\left(  f\right)  }.
\]
So by the union bound, Bob will output $f\left(  x,y\right)  $\ for every
$y\in\mathcal{Y}$\ simultaneously with probability at least%
\[
1-\frac{\left\vert \mathcal{Y}\right\vert }{Q_{2}^{1}\left(  f\right)
^{10}%
}-\frac{1}{Q_{2}^{1}\left(  f\right)  }\geq0.9
\]
for sufficiently large $Q_{2}^{1}\left(  f\right)  $.

Now imagine that the communication channel is blocked, so Bob has to guess
what message Alice wants to send him. \ He does this by using the $K$-qubit
maximally mixed state $I$ in place of $\rho$.\ \ We can write $I$\ as%
\[
I=\frac{1}{2^{K}}\sum_{j=1}^{2^{K}}\left\vert \psi_{j}\right\rangle
\left\langle \psi_{j}\right\vert ,
\]
where $\left\vert \psi_{1}\right\rangle ,\ldots,\left\vert \psi_{2^{K}%
}\right\rangle $\ are orthonormal vectors such that $\left\vert \psi
_{1}\right\rangle =\left\vert \psi\right\rangle $.\ \ So if Bob uses the
same
procedure as above except with $I$ instead of $\rho$, then for any
$\mathcal{Y}\subseteq\mathcal{D}_{x}$ with $\left\vert
\mathcal{Y}\right\vert
\leq Q_{2}^{1}\left(  f\right)  ^{2}$, he will output $f\left(
x,y\right)  $
for every $y\in\mathcal{Y}$ simultaneously with probability at least
$0.9/2^{K}$.

We now give the classical simulation of the quantum protocol. \ Alice's
message to Bob consists of $T\leq K$\ inputs $y_{1},\ldots,y_{T}\in
\mathcal{D}_{x}$, together with $f\left(  x,y_{1}\right)  ,\ldots,f\left(
x,y_{T}\right)  $.\footnote{Strictly speaking, Bob will be able to compute
$f\left(  x,y_{1}\right)  ,\ldots,f\left(  x,y_{T}\right)  $\ for himself
given $y_{1},\ldots,y_{T}$; he does not need Alice to tell him the $f$
values.} \ Thus the message length is $mT+T=O\left(  mQ_{2}^{1}\left(
f\right)  \log Q_{2}^{1}\left(  f\right)  \right)  $. \ Here are the
semantics
of Alice's message: \textit{\textquotedblleft Bob, suppose you looped
over all
}$y\in\mathcal{D}_{x}$\textit{\ in lexicographic order; and for each one,
guessed that }$f\left(  x,y\right)  =\operatorname*{round}\left(
P_{y}\left(
I\right)  \right)  $\textit{, where }$\operatorname*{round}\left(  p\right)
$\textit{ is }$1$\textit{ if }$p\geq1/2$\textit{\ and }$0$\textit{\ if
}$p<1/2$\textit{. \ Then }$y_{1}$\textit{\ is the first }$y$\textit{ for
which
you would guess the wrong value of }$f\left(  x,y\right)  $\textit{. \ In
general, let }$I_{t}$\textit{ be the state obtained by starting from }%
$I$\textit{ and then measuring }$\Lambda\left[  y_{1}\right]  ,\ldots
,\Lambda\left[  y_{t}\right]  $\textit{ in that order, given that the
outcomes
of the measurements are }$f\left(  x,y_{1}\right)  ,\ldots,f\left(
x,y_{t}\right)  $\textit{\ respectively. \ (Note that }$I_{t}$\textit{\
is not
changed by measurements of every }$y\in\mathcal{D}_{x}$\textit{\ up to }%
$y_{t}$\textit{, only by measurements of }$y_{1},\ldots,y_{t}$\textit{.)
\ If
you looped over all }$y\in\mathcal{D}_{x}$\textit{\ in lexicographic order
beginning from }$y_{t}$\textit{,\ then }$y_{t+1}$\textit{ is the first }%
$y$\textit{ you would encounter for which }$\operatorname*{round}\left(
P_{y}\left(  I_{t}\right)  \right)  \neq f\left(  x,y\right)  $%
\textit{.\textquotedblright}

Given the sequence of $y_{t}$'s as defined above, it is obvious that Bob can
compute $f\left(  x,y\right)  $\ for any $y\in\mathcal{D}_{x}$. \ First, if
$y=y_{t}$\ for some $t$, then he simply outputs $f\left(  x,y_{t}\right)  $.
\ Otherwise, let $t^{\ast}$\ be the largest $t$ for which $y_{t}%
<y$\ lexicographically. \ Then Bob prepares a classical description of the
state $I_{t^{\ast}}$---which he can do since he knows $y_{1},\ldots
,y_{t^{\ast}}$\ and $f\left(  x,y_{1}\right)  ,\ldots,f\left(
x,y_{t^{\ast}%
}\right)  $---and then outputs $\operatorname*{round}\left(  P_{y}\left(
I_{t^{\ast}}\right)  \right)  $ as his claimed value of $f\left(  x,y\right)
$. \ Notice that, although Alice uses her knowledge of $\mathcal{D}_{x}$\ to
prepare her message, Bob does not need to know $\mathcal{D}_{x}$\ in
order to
interpret the message. \ That is why the simulation works for partial as
well
as total functions.

But why can we assume that the sequence of $y_{t}$'s stops at $y_{T}$\ for
some $T\leq K$? \ Suppose $T>K$; we will derive a contradiction. \ Let
$\mathcal{Y}=\left\{  y_{1},\ldots,y_{K+1}\right\}  $. \ Then $\left\vert
\mathcal{Y}\right\vert =K+1\leq Q_{2}^{1}\left(  f\right)  ^{2}$, so we know
from previous reasoning that if Bob starts with $I$ and then measures
$\Lambda\left[  y_{1}\right]  ,\ldots,\Lambda\left[  y_{K+1}\right]  $\ in
that order, he will observe $f\left(  x,y_{1}\right)  ,\ldots,f\left(
x,y_{K+1}\right)  $\ simultaneously with probability at least $0.9/2^{K}$.
\ But by the definition of $y_{t}$,\ the probability that $\Lambda\left[
y_{t}\right]  $\ yields the correct outcome is at most $1/2$,\
conditioned on
$\Lambda\left[  y_{1}\right]  ,\ldots,\Lambda\left[  y_{t-1}\right]  $\
having
yielded the correct outcomes. \ Therefore $f\left(  x,y_{1}\right)
,\ldots,f\left(  x,y_{K+1}\right)  $\ are observed simultaneously with
probability at most $1/2^{K+1}<0.9/2^{K}$, contradiction.
\end{proof}

\subsection{Simulating Quantum Advice\label{ADVICESIM}}

We now apply our new simulation method to upper-bound the power of
quantum advice.

\begin{theorem}
\label{qpoly}$\mathsf{BQP/qpoly}\subseteq\mathsf{PP/poly}$.
\end{theorem}

\begin{proof}
For notational convenience, let $L_{n}\left(  x\right)  =1$\ if input
$x\in\left\{  0,1\right\}  ^{n}$\ is in language $L$, and $L_{n}\left(
x\right)  =0$ otherwise. \ Suppose $L_{n}$\ is computed by a $\mathsf{BQP}$
machine using quantum advice of length $p\left(  n\right)  $. \ We will
give a
$\mathsf{PP}$\ machine that computes $L_{n}$ using classical advice of
length
$O\left(  np\left(  n\right)  \log p\left(  n\right)  \right)  $. \
Because of
the close connection between advice and one-way communication, the
simulation
method will be essentially identical to that of %
\autoref{partialthm}.

By using a boosted advice state on $K=O\left(  p\left(  n\right)  \log
p\left(  n\right)  \right)  $\ qubits,\ a polynomial-time quantum algorithm
$A$ can compute $L_{n}\left(  x\right)  $\ with error probability at most
$1/p\left(  n\right)  ^{10}$. \ Now the classical advice to the
$\mathsf{PP}$
machine consists of $T\leq K$\ inputs $x_{1},\ldots,x_{T}\in\left\{
0,1\right\}  ^{n}$, together with $L_{n}\left(  x_{1}\right)  ,\ldots
,L_{n}\left(  x_{T}\right)  $. \ Let $I$ be the maximally mixed state on $K$
qubits. \ Also, let $P_{x}\left(  \rho\right)  $\ be the probability
that $A$
outputs `$1$' on input $x$, given $\rho$\ as its advice state. \ Then
$x_{1}%
$\ is the lexicographically first input $x$\ for which
$\operatorname*{round}%
\left(  P_{x}\left(  I\right)  \right)  \neq L_{n}\left(  x\right)  $. \ In
general, let $I_{t}$\ be the state obtained by starting with $I$ as the
advice
and then running $A$\ on $x_{1},\ldots,x_{t}$\ in that order (uncomputing
garbage along the way), if we postselect on $A$\ correctly outputting
$L_{n}\left(  x_{1}\right)  ,\ldots,L_{n}\left(  x_{t}\right)  $. \ Then
$x_{t+1}$\ is the lexicographically first $x>x_{t}$\ for which
$\operatorname*{round}\left(  P_{x}\left(  I_{t}\right)  \right)  \neq
L_{n}\left(  x\right)  $.

Given the classical advice, we can compute $L_{n}\left(  x\right)  $ as
follows: if $x\in\left\{  x_{1},\ldots,x_{T}\right\}  $\ then output
$L_{n}\left(  x_{t}\right)  $. \ Otherwise let $t^{\ast}$\ be the
largest $t$
for which $x_{t}<x$\ lexicographically,\ and output $\operatorname*{round}%
\left(  P_{x}\left(  I_{t^{\ast}}\right)  \right)  $. \ The proof that this
algorithm works is the same as in %
\autoref{partialthm}, and so is omitted
for brevity. \ All we need to show is that the algorithm can be
implemented in
$\mathsf{PP}$.

Adleman, DeMarrais, and Huang \cite{adh} (see also Fortnow and Rogers
\cite{fr}) showed that $\mathsf{BQP}\subseteq\mathsf{PP}$, by using what
physicists would call a \textquotedblleft Feynman
sum-over-histories.\textquotedblright\ \ Specifically, let $C$ be a
polynomial-size quantum circuit that starts in the all-$0$ state, and that
consists solely of Toffoli and Hadamard gates (Shi \cite{shi:gate}\ has
shown
that this gate set is universal). \ Also, let $\alpha_{z}$\ be the amplitude
of basis state $\left\vert z\right\rangle $ after all gates in $C$ have been
applied. \ We can write $\alpha_{z}$ as a sum of exponentially many
contributions, $a_{1}+\cdots+a_{N}$, where each $a_{i}$\ is a rational real
number computable in classical polynomial time.\ \ So by evaluating the sum%
\[
\left\vert \alpha_{z}\right\vert ^{2}=\sum_{i,j=1}^{N}a_{i}a_{j},
\]
putting positive and negative terms on \textquotedblleft opposite sides
of the
ledger,\textquotedblright\ a $\mathsf{PP}$\ machine can check whether
$\left\vert \alpha_{z}\right\vert ^{2}>\beta$ for any rational constant
$\beta$. \ It follows that a $\mathsf{PP}$\ machine can also check whether%
\[
\sum_{z~:~S_{1}\left(  z\right)  }\left\vert \alpha_{z}\right\vert ^{2}%
 >\sum_{z~:~S_{0}\left(  z\right)  }\left\vert \alpha_{z}\right\vert ^{2}%
\]
(or equivalently, whether $\Pr\left[  S_{1}\right]  >\Pr\left[  S_{0}\right]
$) for any classical polynomial-time predicates $S_{1}$ and $S_{0}$.

Now suppose the circuit $C$ does the following, in the case $x\notin\left\{
x_{1},\ldots,x_{T}\right\}  $. \ It first prepares the $K$-qubit maximally
mixed state $I$ (as half of a $2K$-qubit pure state), and then runs $A$ on
$x_{1},\ldots,x_{t^{\ast}},x$\ in that order, using $I$\ as its advice
state.
\ The claimed values of $L_{n}\left(  x_{1}\right)  ,\ldots,L_{n}\left(
x_{t^{\ast}}\right)  ,L_{n}\left(  x\right)  $\ are written to output
registers but not measured. \ For $i\in\left\{  0,1\right\}  $, let the
predicate $S_{i}\left(  z\right)  $\ hold if and only if basis state
$\left\vert z\right\rangle $\ contains the output sequence $L_{n}\left(
x_{1}\right)  ,\ldots,L_{n}\left(  x_{t^{\ast}}\right)  ,i$. \ Then it
is not
hard to see that%
\[
P_{x}\left(  I_{t^{\ast}}\right)  =\frac{\Pr\left[  S_{1}\right]
}{\Pr\left[
S_{1}\right]  +\Pr\left[  S_{0}\right]  },
\]
so $P_{x}\left(  I_{t^{\ast}}\right)  >1/2$\ and hence $L_{n}\left(
x\right)
=1$\ if and only if $\Pr\left[  S_{1}\right]  >\Pr\left[  S_{0}\right]  $.
\ Since the case $x\in\left\{  x_{1},\ldots,x_{T}\right\}  $\ is
trivial, this
shows that $L_{n}\left(  x\right)  $\ is computable in $\mathsf{PP/poly}$.
\end{proof}

We make five remarks about %
\autoref{qpoly}. \ First, for the same reason
that %
\autoref{partialthm}\ works for partial as well as total functions,
we actually obtain the stronger result that $\mathsf{P{}romiseBQP/qpoly}%
\subseteq\mathsf{P{}romisePP/poly}$, where $\mathsf{P{}romiseBQP}$\ and
$\mathsf{P{}romisePP}$\ are the promise-problem versions of $\mathsf{BQP}%
$\ and $\mathsf{PP}$\ respectively.

Second, as pointed out to us by Lance Fortnow, a corollary of %
\autoref{qpoly} is that we cannot hope to show an unrelativized
separation between
$\mathsf{BQP/poly}$\ and $\mathsf{BQP/qpoly}$, without also showing that
$\mathsf{PP}$\ does not have polynomial-size circuits. \ For
$\mathsf{BQP/poly}\neq\mathsf{BQP/qpoly}$\ clearly implies
that\ $\mathsf{P/poly}\neq\mathsf{PP/poly}$. \ But the latter then implies
that $\mathsf{PP}\not \subset \mathsf{P/poly}$, since assuming $\mathsf{PP}%
\subset\mathsf{P/poly}$\ we could also obtain polynomial-size circuits for a
language $L\in\mathsf{PP/poly}$ by defining a new language $L^{\prime}%
\in\mathsf{PP}$, consisting of all $\left(  x,a\right)  $ pairs such
that the
$\mathsf{PP}$ machine would accept $x$ given advice string $a$. \ The reason
this works is that $\mathsf{PP}$\ is a syntactically defined class.

Third, an earlier version of this paper showed that $\mathsf{BQP/qpoly}%
\subseteq\mathsf{EXP/poly}$, by using a simulation in which an
$\mathsf{EXP}%
$\ machine keeps track of a subspace $H$ of the advice Hilbert space to
which
the `true' advice state must be close. \ In that simulation, the classical
advice specifies inputs $x_{1},\ldots,x_{T}$\ for which $\dim\left(
H\right)
$\ is at least halved; the observation that $\dim\left(  H\right)  $\
must be
at least $1$ by the end then implies that $T\leq K=O\left(  p\left(
n\right)
\log p\left(  n\right)  \right)  $, meaning that the advice is of polynomial
size. \ The huge improvement from\ $\mathsf{EXP}$\ to $\mathsf{PP}$\ came
solely from working with \textit{measurement outcomes} and their
\textit{probabilities} instead of with \textit{subspaces} and their
\textit{dimensions}. \ We can compute the former using the same
\textquotedblleft Feynman sum-over-histories\textquotedblright\ that Adleman
et al. \cite{adh}\ used to show $\mathsf{BQP}\subseteq\mathsf{PP}$, but
could
not see any way to compute the latter without explicitly storing and
diagonalizing exponentially large matrices.

Fourth, assuming $\mathsf{BQP/poly}\neq\mathsf{BQP/qpoly}$, %
\autoref{qpoly}\ is \textit{almost} the best result of its kind that one
could
hope for, since the only classes known to lie between $\mathsf{BQP}$\ and
$\mathsf{PP}$\ and not known to equal either are obscure ones such as
$\mathsf{AWPP}$\ \cite{fr}. \ Initially the theorem\ seemed to us to prove
something stronger, namely that $\mathsf{BQP/qpoly}\subseteq
\mathsf{PostBQP/poly}$. \ Here $\mathsf{PostBQP}$\ is the class of languages
decidable by polynomial-size quantum circuits with \textit{postselection}%
---meaning the ability to measure a qubit that has a nonzero probability of
being $\left\vert 1\right\rangle $, and then \textit{assume} that
the measurement outcome will be $\left\vert 1\right\rangle $. \
Clearly $\mathsf{PostBQP}$\ lies somewhere between $\mathsf{BQP}$\
and $\mathsf{PP}$; one can think of it as a quantum analogue of the
classical complexity class $\mathsf{BPP}_{\mathsf{path}}$\
\cite{hht}. \ We have since shown, however, that
$\mathsf{PostBQP}=\mathsf{PP}$ \cite{aar:pp}.

Fifth, it is clear that Adleman et al.'s $\mathsf{BQP}\subseteq\mathsf{PP}%
$\ result \cite{adh}\ can be extended to show that
$\mathsf{PQP}=\mathsf{PP}$.
\ Here $\mathsf{PQP}$\ is the quantum analogue of $\mathsf{PP}$---that is,
quantum polynomial time but where the probability of a correct answer need
only be bounded above $1/2$, rather than above $2/3$. \ A reviewer asked
whether %
\autoref{qpoly} could similarly be extended to show that
$\mathsf{PQP/qpoly}=\mathsf{PP/poly}$. \ The answer is no---for indeed,
$\mathsf{PQP/qpoly}$\ contains every language whatsoever! \ To see this,
given
any function $L_{n}:\left\{  0,1\right\}  ^{n}\rightarrow\left\{
0,1\right\}
$, let our quantum advice state be%
\[
\left\vert \psi_{n}\right\rangle =\frac{1}{2^{n/2}}\sum_{x\in\left\{
0,1\right\}  ^{n}}\left\vert x\right\rangle \left\vert L_{n}\left(  x\right)
\right\rangle .
\]
Then a $\mathsf{PQP}$\ algorithm to compute $L_{n}$ is as follows: given an
input $x\in\left\{  0,1\right\}  ^{n}$, first measure $\left\vert \psi
_{n}\right\rangle $ in the standard basis. \ If $\left\vert x\right\rangle
\left\vert L_{n}\left(  x\right)  \right\rangle $\ is observed, output
$L_{n}\left(  x\right)  $; otherwise output a uniform random bit.

\section{Oracle Limitations\label{ORACLE}}

Can quantum computers solve $\mathsf{NP}$-complete problems\ in polynomial
time?\ \ In the early days of quantum computing, Bennett et al. \cite{bbbv}%
\ gave an oracle relative to which $\mathsf{NP}\not \subset \mathsf{BQP}$,
providing what is still the best evidence we have that the answer is no.
\ It
is easy to extend Bennett et al.'s result to give an oracle relative to
which
$\mathsf{NP}\not \subset \mathsf{BQP/poly}$; that is, $\mathsf{NP}$ is hard
even for nonuniform quantum algorithms. \ But when we try to show
$\mathsf{NP}\not \subset \mathsf{BQP/qpoly}$\ relative to an oracle, a new
difficulty arises: even if the oracle encodes $2^{n}$ exponentially hard
search problems for each input length $n$, the quantum advice, being an
\textquotedblleft exponentially large object\textquotedblright\ itself,
might
somehow encode information about all $2^{n}$ problems. \ We need to
argue that
even if so, only a miniscule fraction of that information can be
extracted by
measuring the advice.

How does one prove such a statement? \ As it turns out, the task can be
reduced to proving a \textit{direct product theorem} for quantum search.
\ This is a theorem that in its weakest form says the following: given $N$
items, $K$ of which are marked, if we lack enough time to find even
\textit{one} marked item, then the probability of finding all $K$ items
decreases exponentially in $K$. \ For intuitively, suppose there were a
quantum advice state that let us efficiently find any one of $K$ marked
items.
\ Then by \textquotedblleft guessing\textquotedblright\ the advice (i.e.
replacing it by a maximally mixed state), and then using the guessed advice
multiple times, we could efficiently find all $K$ of the items with a
success
probability that our direct product theorem shows is impossible. \ This
reduction is formalized in %
\autoref{npqpoly}.

But what about the direct product theorem itself? \ It seems like it
should be trivial to prove---for surely there are no devious
correlations by which success in finding one marked item leads to
success in finding all the others! \ So it is surprising that even a
weak direct product theorem eluded proof for years. \ In 2001,
Klauck \cite{klauck:ts}\ gave an attempted proof using the hybrid
method of Bennett et al. \cite{bbbv}. \ His motivation was to show a
limitation of space-bounded quantum sorting algorithms. \
Unfortunately, Klauck's proof contained a
bug.\footnote{Specifically, the last sentence in the proof of Lemma
5 in \cite{klauck:ts}\ (\textquotedblleft Clearly this probability
is at least $q_{x}\left( p_{x}-\alpha\right)
$\textquotedblright%
) is not justified by what precedes it.}

In this section we give the first correct proof of a direct product
theorem, based on the polynomial method of Beals et al.
\cite{bbcmw}. \ Besides showing that $\mathsf{NP}\not \subset
\mathsf{BQP/qpoly}$\ relative to an oracle, our result can be used
to recover the claims made in \cite{klauck:ts} about the hardness of
quantum sorting (see Klauck, \v{S}palek, and de Wolf \cite{ksw}\ for
details). \ We expect the result to have other applications as well.

We will need the following lemma of Beals et al. \cite{bbcmw},\ which builds
on ideas due to Minsky and Papert \cite{mp}\ and Nisan and Szegedy
\cite{ns}.

\begin{lemma}
[Beals et al.]\label{bbcmwlem}Suppose a quantum algorithm makes $T$
queries to
an oracle string $X\in\left\{  0,1\right\}  ^{N}$, and accepts with
probability $A\left(  X\right)  $. \ Then there exists a real polynomial
$p$,
of degree at most $2T$, such that%
\[
p\left(  i\right)  =\operatorname*{EX}_{\left\vert X\right\vert =i}\left[
A\left(  X\right)  \right]
\]
for all integers $i\in\left\{  0,\ldots,N\right\}  $, where $\left\vert
X\right\vert $\ denotes the Hamming weight of $X$.
\end{lemma}

\lemref{bbcmwlem}\ implies that, to lower-bound the number of queries $T$
made by a quantum algorithm, it suffices to lower-bound $\deg\left(
p\right)
$, where $p$ is a real polynomial representing the algorithm's expected
acceptance probability. \ As an example, any quantum algorithm that computes
the $\operatorname*{OR}$\ function on $N$ bits, with success probability at
least $2/3$, yields a polynomial $p$\ such that $p\left(  0\right)
\in\left[
0,1/3\right]  $\ and $p\left(  i\right)  \in\left[  2/3,1\right]  $\ for all
integers $i\in\left\{  1,\ldots,N\right\}  $. \ To lower-bound the degree of
such a polynomial, one can use an inequality proved by A. A. Markov in 1890
(\cite{aamarkov}; see also \cite{rivlin}):

\begin{theorem}
[A. A. Markov]\label{aamarkovthm}Given a real polynomial $p$ and constant
$N>0$, let $r^{\left(  0\right)  }=\max_{x\in\left[  0,N\right]  }\left\vert
p\left(  x\right)  \right\vert $\ and $r^{\left(  1\right)  }=\max
_{x\in\left[  0,N\right]  }\left\vert p^{\prime}\left(  x\right)
\right\vert
$. \ Then%
\[
\deg\left(  p\right)  \geq\sqrt{\frac{Nr^{\left(  1\right)  }}{2r^{\left(
0\right)  }}}.
\]

\end{theorem}

\autoref{aamarkovthm} deals with the entire range $\left[  0,N\right]  $,
whereas in our setting $p\left(  x\right)  $\ is constrained only at the
integer points $x\in\left\{  0,\ldots,N\right\}  $. \ But as shown in
\cite{ez,ns,rc}, this is not a problem. \ For by elementary calculus,
$p\left(  0\right)  \leq1/3$\ and $p\left(  1\right)  \geq2/3$\ imply that
$p^{\prime}\left(  x\right)  \geq1/3$\ for some real $x\in\left[  0,1\right]
$, and therefore $r^{\left(  1\right)  }\geq1/3$. \ Furthermore, let
$x^{\ast
}$\ be a point in $\left[  0,N\right]  $\ where $\left\vert p\left(  x^{\ast
}\right)  \right\vert =r^{\left(  0\right)  }$. \ Then $p\left(
\left\lfloor
x^{\ast}\right\rfloor \right)  \in\left[  0,1\right]  $\ and $p\left(
\left\lceil x^{\ast}\right\rceil \right)  \in\left[  0,1\right]  $\
imply that
$r^{\left(  1\right)  }\geq2\left(  r^{\left(  0\right)  }-1\right)  $.
\ Thus%
\[
\deg\left(  p\right)  \geq\sqrt{\frac{Nr^{\left(  1\right)  }}{2r^{\left(
0\right)  }}}\geq\sqrt{\frac{N\max\left\{  1/3,2\left(  r^{\left(  0\right)
}-1\right)  \right\}  }{2r^{\left(  0\right)  }}}=\Omega\left(  \sqrt
{N}\right)  .
\]
This is the proof of\ Beals et al. \cite{bbcmw}\ that quantum search
requires
$\Omega\left(  \sqrt{N}\right)  $\ queries.

When proving a direct product theorem, we can no longer apply %
\autoref{aamarkovthm} so straightforwardly. \ The reason is that the success
probabilities in question are extremely small, and therefore the maximum
derivative $r^{\left(  1\right)  }$\ could also be extremely small.
\ Fortunately, though, we can still prove a good lower bound on the
degree of
the relevant polynomial $p$. \ The key is to look not just at the first
derivative of $p$, but at higher derivatives.

To start, we need a lemma about the behavior of functions under repeated
differentiation.

\begin{lemma}
\label{derivative}Let $f:\mathbb{R}\rightarrow\mathbb{R}$\ be an infinitely
differentiable function such that for some positive integer $K$, we have
$f\left(  i\right)  =0$ for all $i\in\left\{  0,\ldots,K-1\right\}  $ and
$f\left(  K\right)  =\delta>0$. \ Also, let $r^{\left(  m\right)  }=\max
_{x\in\left[  0,N\right]  }\left\vert f^{\left(  m\right)  }\left(  x\right)
\right\vert $, where $f^{\left(  m\right)  }\left(  x\right)  $\ is the
$m^{th}$\ derivative of $f$ evaluated at $x$ (thus $f^{\left(  0\right)  }%
=f$). \ Then $r^{\left(  m\right)  }\geq\delta/m!$\ for all $m\in\left\{
0,\ldots,K\right\}  $.
\end{lemma}

\begin{proof}
We claim, by induction on $m$, that there exist $K-m+1$\ points $0\leq
x_{0}^{\left(  m\right)  }<\cdots<x_{K-m}^{\left(  m\right)  }\leq K$ such
that $f^{\left(  m\right)  }\left(  x_{i}^{\left(  m\right)  }\right)
=0$ for
all $i\leq K-m-1$\ and $f^{\left(  m\right)  }\left(  x_{K-m}^{\left(
m\right)  }\right)  \geq\delta/m!$. \ If we define $x_{i}^{\left(  0\right)
}=i$, then the base case $m=0$\ is immediate from the conditions of the
lemma.
\ Suppose the claim is true for $m$; then by elementary calculus, for all
$i\leq K-m-2$\ there exists a point $x_{i}^{\left(  m+1\right)  }\in\left(
x_{i}^{\left(  m\right)  },x_{i+1}^{\left(  m\right)  }\right)  $\ such that
$f^{\left(  m+1\right)  }\left(  x_{i}^{\left(  m+1\right)  }\right)  =0$.
\ Notice that $x_{i}^{\left(  m+1\right)  }\geq x_{i}^{\left(  m\right)  }%
\geq\cdots\geq x_{i}^{\left(  0\right)  }=i$. \ So there is also a point
$x_{K-m-1}^{\left(  m+1\right)  }\in\left(  x_{K-m-1}^{\left(  m\right)
},x_{K-m}^{\left(  m\right)  }\right)  $ such that\
\begin{align*}
f^{\left(  m+1\right)  }\left(  x_{K-m-1}^{\left(  m+1\right)  }\right)   &
\geq\frac{f^{\left(  m\right)  }\left(  x_{K-m}^{\left(  m\right)  }\right)
-f^{\left(  m\right)  }\left(  x_{K-m-1}^{\left(  m\right)  }\right)
}{x_{K-m}^{\left(  m\right)  }-x_{K-m-1}^{\left(  m\right)  }}\\
&  \geq\frac{\delta/m!-0}{K-\left(  K-m-1\right)  }\\
&  =\frac{\delta}{\left(  m+1\right)  !}.
\end{align*}

\end{proof}

With the help of %
\lemref{derivative}, we can sometimes lower-bound the
degree of a real polynomial even its first derivative is small
throughout the
region of interest.\ To do so, we use the following generalization of A. A.
Markov's inequality (%
\autoref{aamarkovthm}), which was\ proved by A. A.
Markov's younger brother V. A. Markov\ in 1892 (\cite{vamarkov}; see also
\cite{rivlin}).

\begin{theorem}
[V. A. Markov]\label{vamarkovthm}Given a real polynomial $p$ of degree
$d$ and
positive real number $N$, let $r^{\left(  m\right)  }=\max_{x\in\left[
0,N\right]  }\left\vert p^{\left(  m\right)  }\left(  x\right)
\right\vert $.
\ Then for all $m\in\left\{  1,\ldots,d\right\}  $,%
\begin{align*}
r^{\left(  m\right)  }  &  \leq\left(  \frac{2r^{\left(  0\right)  }}%
{N}\right)  ^{m}T_{d}^{\left(  m\right)  }\left(  1\right) \\
&  \leq\left(  \frac{2r^{\left(  0\right)  }}{N}\right)  ^{m}\frac
{d^{2}\left(  d^{2}-1^{2}\right)  \left(  d^{2}-2^{2}\right)  \cdot\cdots
\cdot\left(  d^{2}-\left(  m-1\right)  ^{2}\right)  }{1\cdot3\cdot5\cdot
\cdots\cdot\left(  2m-1\right)  }.
\end{align*}
Here $T_{d}\left(  x\right)  =\cos\left(  d\arccos x\right)  $\ is the
$d^{th}$\ Chebyshev polynomial of the first kind.
\end{theorem}

As we demonstrate\ below, combining %
\autoref{vamarkovthm}\ with %
\lemref{derivative} yields a lower bound on $\deg\left(  p\right)  $.

\begin{lemma}
\label{degreelb}Let $p$\ be a real polynomial such that

\begin{enumerate}
\item[(i)] $p\left(  x\right)  \in\left[  0,1\right]  $ at all integer
points
$x\in\left\{  0,\ldots,N\right\}  $, and

\item[(ii)] for some positive integer $K\leq N$ and real $\delta>0$, we have
$p\left(  K\right)  =\delta$\ and $p\left(  i\right)  =0$\ for all
$i\in\left\{  0,\ldots,K-1\right\}  $.
\end{enumerate}

Then $\deg\left(  p\right)  =\Omega\left(  \sqrt{N\delta^{1/K}}\right)  $.
\end{lemma}

\begin{proof}
Let $p^{\left(  m\right)  }$\ and $r^{\left(  m\right)  }$ be as in %
\autoref{vamarkovthm}.\ \ Then for all $m\in\left\{  1,\ldots,\deg\left(
p\right)  \right\}  $, %
\autoref{vamarkovthm}\ yields%
\[
r^{\left(  m\right)  }\leq\left(  \frac{2r^{\left(  0\right)  }}{N}\right)
^{m}\frac{\deg\left(  p\right)  ^{2m}}{1\cdot3\cdot5\cdot\cdots\cdot\left(
2m-1\right)  }%
\]
Rearranging,%
\[
\deg\left(  p\right)  \geq\sqrt{\frac{N}{2r^{\left(  0\right)  }}\left(
1\cdot3\cdot5\cdot\cdots\cdot\left(  2m-1\right)  \cdot r^{\left(  m\right)
}\right)  ^{1/m}}%
\]
for all $m\geq1$\ (if $m>\deg\left(  p\right)  $\ then $r^{\left(  m\right)
}=0$\ so the bound is trivial).

There are now two cases. \ First suppose $r^{\left(  0\right)  }\geq
2$.\ \ Then as discussed previously, condition (i) implies that $r^{\left(
1\right)  }\geq2\left(  r^{\left(  0\right)  }-1\right)  $, and hence that%
\[
\deg\left(  p\right)  \geq\sqrt{\frac{Nr^{\left(  1\right)  }}{2r^{\left(
0\right)  }}}\geq\sqrt{\frac{N\left(  r^{\left(  0\right)  }-1\right)
}{r^{\left(  0\right)  }}}=\Omega\left(  \sqrt{N}\right)
\]
by %
\autoref{aamarkovthm}. \ Next suppose $r^{\left(  0\right)  }<2$.
\ Then $r^{\left(  m\right)  }\geq\delta/m!$\ for all $m\leq K$\ by %
\lemref{derivative}. \ So setting $m=K$\ yields%
\[
\deg\left(  p\right)  \geq\sqrt{\frac{N}{4}\left(  1\cdot3\cdot5\cdot
\cdots\cdot\left(  2K-1\right)  \cdot\frac{\delta}{K!}\right)  ^{1/K}}%
=\Omega\left(  \sqrt{N\delta^{1/K}}\right)  .
\]
Either way we are done.
\end{proof}

Strictly speaking, we do not need the full strength of %
\autoref{vamarkovthm}\ to prove a lower bound on $\deg\left(  p\right)
$ that
suffices for an oracle separation between $\mathsf{NP}$\ and
$\mathsf{BQP/qpoly}$. \ For we can show a \textquotedblleft
rough-and-ready\textquotedblright\ version of V. A. Markov's inequality by
applying A. A. Markov's inequality (%
\autoref{aamarkovthm}) repeatedly, to
$p,p^{\left(  1\right)  },p^{\left(  2\right)  },$ and so on. \ This yields%
\[
r^{\left(  m\right)  }\leq\frac{2}{N}\deg\left(  p\right)  ^{2}r^{\left(
m-1\right)  }\leq\left(  \frac{2}{N}\deg\left(  p\right)  ^{2}\right)
^{m}r^{\left(  0\right)  }%
\]
\begin{sloppypar}
\noindent
for all $m$. \ If $\deg\left(  p\right)  $\ is small, then this upper
bound on
$r^{\left(  m\right)  }$\ contradicts the lower bound of %
\lemref{derivative}. \ However, the lower bound on $\deg\left(
p\right)  $\ that
one gets from A. A. Markov's inequality is only $\Omega\left(  \sqrt
{N\delta^{1/K}/K}\right)  $, as opposed to $\Omega\left(
\sqrt{N\delta^{1/K}%
}\right)  $\ from %
\lemref{degreelb}.\footnote{An earlier version of this
paper claimed to prove $\deg\left(  p\right)  =\Omega\left(  \sqrt{NK}%
/\log^{3/2}\left(  1/\delta\right)  \right)  $, by applying
\textit{Bernstein's inequality} \cite{bernstein}\ rather than A. A. Markov's
to all derivatives $p^{\left(  m\right)  }$. \ We have since discovered
a flaw
in that argument. \ In any case, the Bernstein\ lower bound is both
unnecessary for an oracle separation, and superseded by the later results of
Klauck et al. \cite{ksw}.}
\end{sloppypar}

Shortly after seeing our proof of a weak direct product theorem, Klauck,
\v{S}palek, and de Wolf \cite{ksw}\ managed to improve the lower bound on
$\deg\left(  p\right)  $\ to the essentially tight $\Omega\left(
\sqrt{NK\delta^{1/K}}\right)  $. \ In particular, their bound implies that
$\delta$\ decreases exponentially in $K$ whenever $\deg\left(  p\right)
=o\left(  \sqrt{NK}\right)  $. \ They obtained this improvement by
\textit{factoring} $p$ instead of differentiating it as in %
\lemref{derivative}.

In any case, a direct product theorem follows trivially from what has
already
been said.

\begin{theorem}
[Direct Product Theorem]\label{directprod}Suppose a quantum algorithm makes
$T$ queries to an oracle string $X\in\left\{  0,1\right\}  ^{N}$. \ Let
$\delta$\ be the minimum probability, over all $X$\ with Hamming weight
$\left\vert X\right\vert =K$, that the algorithm finds all $K$\ of the
`$1$'\ bits. \ Then $\delta\leq\left(  cT^{2}/N\right)  ^{K}$ for some
constant $c$.
\end{theorem}

\begin{proof}
Have the algorithm accept if it finds $K$ or more `$1$'\ bits and reject
otherwise. \ Let $p\left(  i\right)  $\ be the\ expected probability of
acceptance if $X$\ is drawn uniformly at random subject to $\left\vert
X\right\vert =i$. \ Then we know the following about $p$:

\begin{enumerate}
\item[(i)] $p\left(  i\right)  \in\left[  0,1\right]  $ at all integer
points
$i\in\left\{  0,\ldots,N\right\}  $, since $p\left(  i\right)  $\ is a
probability.

\item[(ii)] $p\left(  i\right)  =0$ for all $i\in\left\{  0,\ldots
,K-1\right\}  $, since there are not $K$ marked items to be found.

\item[(iii)] $p\left(  K\right)  \geq\delta$.
\end{enumerate}

Furthermore, %
\lemref{bbcmwlem} implies that $p$\ is a polynomial in $i$
satisfying $\deg\left(  p\right)  \leq2T$. \ It follows from %
\lemref{degreelb} that $T=\Omega\left(  \sqrt{N\delta^{1/K}}\right)  $, or
rearranging, that $\delta\leq\left(  cT^{2}/N\right)  ^{K}$ for some
constant
$c$.
\end{proof}

We can now prove the desired oracle separation using standard complexity
theory tricks.

\begin{theorem}
\label{npqpoly}There exists an oracle relative to which $\mathsf{NP}%
\not \subset \mathsf{BQP/qpoly}$.
\end{theorem}
\eject

\begin{proof}
Given an oracle $A:\left\{  0,1\right\}  ^{\ast}\rightarrow\left\{
0,1\right\}  $, define the language $L_{A}$\ by $\left(  y,z\right)  \in
L_{A}$\ if and only if $y\leq z$\ lexicographically and there exists an $x$
such that $y\leq x\leq z$\ and $A\left(  x\right)  =1$. \ Clearly $L_{A}%
\in\mathsf{NP}^{A}$\ for all $A$. \ We argue that for some $A$, no
$\mathsf{BQP/qpoly}$\ machine $M$ with oracle access to $A$ can decide
$L_{A}%
$. \ Without loss of generality we assume $M$ is fixed, so that only the
advice states $\left\{  \left\vert \psi_{n}\right\rangle \right\}
_{n\geq1}%
$\ depend on $A$. \ We also assume the advice is boosted, so that $M$'s
error
probability on any input $\left(  y,z\right)  $ is $2^{-\Omega\left(
n^{2}\right)  }$.

Choose a set $S\subset\left\{  0,1\right\}  ^{n}$ subject to $\left\vert
S\right\vert =2^{n/10}$; then for all $x\in\left\{  0,1\right\}  ^{n}$, set
$A\left(  x\right)  =1$\ if and only if $x\in S$. \ We claim that by using
$M$, an algorithm could find all $2^{n/10}$\ elements of $S$\ with high
probability after only $2^{n/10}\operatorname*{poly}\left(  n\right)
$\ queries to $A$. \ Here is how: first use binary search (repeatedly
halving
the distance between $y$\ and $z$) to find the lexicographically first
element
of $S$.\ \ By %
\lemref{tracedist}, the boosted advice state $\left\vert
\psi_{n}\right\rangle $\ is good for $2^{\Omega\left(  n^{2}\right)  }$\
uses,
so this takes only $\operatorname*{poly}\left(  n\right)  $\ queries. \ Then
use binary search to find the lexicographically second element, and so on
until all\ elements have been found.

Now replace $\left\vert \psi_{n}\right\rangle $\ by the maximally mixed
state
as in %
\autoref{partialthm}. This yields an algorithm that uses no advice,
makes $2^{n/10}\operatorname*{poly}\left(  n\right)  $ queries, and
finds all
$2^{n/10}$\ elements of\ $S$\ with probability $2^{-O\left(
\operatorname*{poly}\left(  n\right)  \right)  }$. \ But taking $\delta
=2^{-O\left(  \operatorname*{poly}\left(  n\right)  \right)  }$,
$T=2^{n/10}\operatorname*{poly}\left(  n\right)  $, $N=2^{n}$, and
$K=2^{n/10}$, such an algorithm would satisfy $\delta\gg\left(  cT^{2}%
/N\right)  ^{K}$, which violates the bound of %
\autoref{directprod}.
\end{proof}

Indeed one can show that $\mathsf{NP}\not \subset \mathsf{BQP/qpoly}%
$\ relative a random oracle with probability $1$.\footnote{First group the
oracle bits into polynomial-size blocks as Bennett and Gill \cite{bg}
do, then
use the techniques of Aaronson \cite{aar:col} to show that the acceptance
probability is a low-degree univariate polynomial in the number of all-$0$
blocks. \ The rest of the proof follows %
\autoref{npqpoly}.}

\section{The Trace Distance Method\label{TDIST}}

This section introduces a new method for proving lower bounds on quantum
one-way communication complexity. \ Unlike in %
\secref{1WAYSIM}, here we
do not try to simulate quantum protocols using classical ones. \ Instead we
prove lower bounds for quantum protocols directly, by reasoning about the
trace distance between two possible distributions over Alice's quantum
message
(that is, between two mixed states). \ The result is a method that works
even
if Alice's and Bob's inputs are the same size.

We first state our method as a general theorem; then, in %
\secref{APPL},
we apply the theorem to prove lower bounds for two problems of Ambainis.
\ Let
$\left\Vert \mathcal{D}-\mathcal{E}\right\Vert $\ denote the variation
distance between probability distributions $\mathcal{D}$ and $\mathcal{E}$.

\begin{theorem}
\label{vardist}Let $f:\left\{  0,1\right\}  ^{n}\times\left\{  0,1\right\}
^{m}\rightarrow\left\{  0,1\right\}  $\ be a total Boolean function. \ For
each $y\in\left\{  0,1\right\}  ^{m}$, let $\mathcal{A}_{y}$\ be a
distribution over $x\in\left\{  0,1\right\}  ^{n}$ such that $f\left(
x,y\right)  =1$. \ Let $\mathcal{B}$\ be a distribution over $y\in\left\{
0,1\right\}  ^{m}$, and let $\mathcal{D}_{k}$ be the distribution over
$\left(  \left\{  0,1\right\}  ^{n}\right)  ^{k}$\ formed by first choosing
$y\in\mathcal{B}$\ and then choosing $k$ samples independently from
$\mathcal{A}_{y}$. \ Suppose that $\Pr_{x\in\mathcal{D}_{1},y\in\mathcal{B}%
}\left[  f\left(  x,y\right)  =0\right]  =\Omega\left(  1\right)  $\ and
that
$\left\Vert \mathcal{D}_{2}-\mathcal{D}_{1}^{2}\right\Vert \leq\delta.$
\ Then
$Q_{2}^{1}\left(  f\right)  =\Omega\left(  \log1/\delta\right)  $.
\end{theorem}

\begin{proof}
Suppose that if Alice's input is $x$, then she sends Bob the $l$-qubit mixed
state $\rho_{x}$. \ Suppose also that for every $x\in\left\{  0,1\right\}
^{n}$ and $y\in\left\{  0,1\right\}  ^{m}$, Bob outputs $f\left(  x,y\right)
$\ with probability at least $2/3$. \ Then by amplifying a constant
number of
times, Bob's success probability can be made $1-\varepsilon$\ for any
constant
$\varepsilon>0$. \ So with $L=O\left(  l\right)  $\ qubits of communication,
Bob can distinguish the following two cases with constant bias:

\textbf{Case I.} $\ y$ was drawn from $\mathcal{B}$\ and $x$ from
$\mathcal{D}_{1}$.

\textbf{Case II.} $y$ was drawn from $\mathcal{B}$\ and $x$ from
$\mathcal{A}_{y}$.

For in Case I, we assumed that $f\left(  x,y\right)  =0$\ with constant
probability, whereas in Case II, $f\left(  x,y\right)  =1$\ always. \ An
equivalent way to say this is that with constant probability over $y$,
Bob can
distinguish the mixed states $\rho=\operatorname*{EX}_{x\in\mathcal{D}_{1}%
}\left[  \rho_{x}\right]  $\ and $\rho_{y}=\operatorname*{EX}_{x\in
\mathcal{A}_{y}}\left[  \rho_{x}\right]  $\ with constant bias. \ Therefore%
\[
\operatorname*{EX}_{y\in\mathcal{B}}\left[  \left\Vert \rho-\rho
_{y}\right\Vert _{\operatorname*{tr}}\right]  =\Omega\left(  1\right)  .
\]

We need an upper bound on the trace distance $\left\Vert \rho-\rho
_{y}\right\Vert _{\operatorname*{tr}}$\ that is more amenable to analysis.
\ Let $\lambda_{1},\ldots,\lambda_{2^{L}}$\ be the eigenvalues of $\rho
-\rho_{y}$. \ Then%
\begin{align*}
\left\Vert \rho-\rho_{y}\right\Vert _{\operatorname*{tr}}  &  =\frac{1}{2}%
\sum_{i=1}^{2^{L}}\left\vert \lambda_{i}\right\vert \\
&  \leq\frac{1}{2}\sqrt{2^{L}\sum_{i=1}^{2^{L}}\lambda_{i}^{2}}\\
&  =2^{L/2-1}\sqrt{\sum_{i,j=1}^{2^{L}}\left\vert \left(  \rho\right)
_{ij}-\left(  \rho_{y}\right)  _{ij}\right\vert ^{2}}%
\end{align*}
where $\left(  \rho\right)  _{ij}$\ is the $\left(  i,j\right)  $\ entry of
$\rho$. \ Here the second line uses the Cauchy-Schwarz inequality, and the
third line uses the unitary invariance of the Frobenius norm.

We claim that%
\[
\operatorname*{EX}_{y\in\mathcal{B}}\left[  \sum_{i,j=1}^{2^{L}}\left\vert
\left(  \rho\right)  _{ij}-\left(  \rho_{y}\right)  _{ij}\right\vert
^{2}\right]  \leq2\delta.
\]
{}From this claim it follows that%
\begin{align*}
\operatorname*{EX}_{y\in\mathcal{B}}\left[  \left\Vert \rho-\rho
_{y}\right\Vert _{\operatorname*{tr}}\right]   &  \leq2^{L/2-1}%
\operatorname*{EX}_{y\in\mathcal{B}}\left[  \sqrt{\sum_{i,j=1}^{2^{L}%
}\left\vert \left(  \rho\right)  _{ij}-\left(  \rho_{y}\right)  _{ij}%
\right\vert ^{2}}\right] \\
&  \leq2^{L/2-1}\sqrt{\operatorname*{EX}_{y\in\mathcal{B}}\left[  \sum
_{i,j=1}^{2^{L}}\left\vert \left(  \rho\right)  _{ij}-\left(
\rho_{y}\right)
_{ij}\right\vert ^{2}\right]  }\\
&  \leq\sqrt{2^{L-1}\delta}.
\end{align*}
Therefore the message length $L$ must be $\Omega\left(  \log1/\delta\right)
$\ to ensure that $\operatorname*{EX}_{y\in\mathcal{B}}\left[  \left\Vert
\rho-\rho_{y}\right\Vert _{\operatorname*{tr}}\right]  =\Omega\left(
1\right)  $.

Let us now prove the claim. \ We have%
\begin{align*}
\operatorname*{EX}_{y\in\mathcal{B}}\left[  \sum_{i,j=1}^{2^{L}}\left\vert
\left(  \rho\right)  _{ij}-\left(  \rho_{y}\right)  _{ij}\right\vert
^{2}\right]   &  =\sum_{i,j=1}^{2^{L}}\left(  \left\vert \left(  \rho\right)
_{ij}\right\vert ^{2}-2\operatorname{Re}\left(  \left(  \rho\right)
_{ij}^{\ast}\operatorname*{EX}_{y\in\mathcal{B}}\left[  \left(  \rho
_{y}\right)  _{ij}\right]  \right)  +\operatorname*{EX}_{y\in\mathcal{B}%
}\left[  \left\vert \left(  \rho_{y}\right)  _{ij}\right\vert ^{2}\right]
\right) \\
&  =\sum_{i,j=1}^{2^{L}}\left(  \operatorname*{EX}_{y\in\mathcal{B}}\left[
\left\vert \left(  \rho_{y}\right)  _{ij}\right\vert ^{2}\right]
-\left\vert
\left(  \rho\right)  _{ij}\right\vert ^{2}\right)  ,
\end{align*}
since $\operatorname*{EX}_{y\in\mathcal{B}}\left[  \left(  \rho_{y}\right)
_{ij}\right]  =\left(  \rho\right)  _{ij}$. \ For a given $\left(
i,j\right)
$\ pair,%
\begin{align*}
\operatorname*{EX}_{y\in\mathcal{B}}\left[  \left\vert \left(  \rho
_{y}\right)  _{ij}\right\vert ^{2}\right]  -\left\vert \left(  \rho\right)
_{ij}\right\vert ^{2}  &  =\operatorname*{EX}_{y\in\mathcal{B}}\left[
\left\vert \operatorname*{EX}_{x\in\mathcal{A}_{y}}\left[  \left(  \rho
_{x}\right)  _{ij}\right]  \right\vert ^{2}\right]  -\left\vert
\operatorname*{EX}_{x\in\mathcal{D}_{1}}\left[  \left(  \rho_{x}\right)
_{ij}\right]  \right\vert ^{2}\\
&  =\operatorname*{EX}_{y\in\mathcal{B},x,z\in\mathcal{A}_{y}}\left[  \left(
\rho_{x}\right)  _{ij}^{\ast}\left(  \rho_{z}\right)  _{ij}\right]
-\operatorname*{EX}_{x,z\in\mathcal{D}_{1}}\left[  \left(  \rho_{x}\right)
_{ij}^{\ast}\left(  \rho_{z}\right)  _{ij}\right] \\
&  =\sum_{x,z}\left(  \Pr_{\mathcal{D}_{2}}\left[  x,z\right]  -\Pr
_{\mathcal{D}_{1}^{2}}\left[  x,z\right]  \right)  \left(  \rho_{x}\right)
_{ij}^{\ast}\left(  \rho_{z}\right)  _{ij}.
\end{align*}
Now for all $x,z$,%
\[
\left\vert \sum_{i,j=1}^{2^{L}}\left(  \rho_{x}\right)  _{ij}^{\ast}\left(
\rho_{z}\right)  _{ij}\right\vert \leq\sum_{i,j=1}^{2^{L}}\left\vert \left(
\rho_{x}\right)  _{ij}\right\vert ^{2}\leq1.
\]
Hence%
\begin{align*}
\sum_{x,z}\left(  \Pr_{\mathcal{D}_{2}}\left[  x,z\right]
-\Pr_{\mathcal{D}%
_{1}^{2}}\left[  x,z\right]  \right)  \sum_{i,j=1}^{2^{L}}\left(  \rho
_{x}\right)  _{ij}^{\ast}\left(  \rho_{z}\right)  _{ij}  &  \leq\sum
_{x,z}\left(  \Pr_{\mathcal{D}_{2}}\left[  x,z\right]  -\Pr_{\mathcal{D}%
_{1}^{2}}\left[  x,z\right]  \right) \\
&  =2\left\Vert \mathcal{D}_{2}-\mathcal{D}_{1}^{2}\right\Vert \\
&  \leq2\delta,
\end{align*}
and we are done.
\end{proof}

The difficulty in extending %
\autoref{vardist}\ to partial functions is
that the distribution $\mathcal{D}_{1}$\ might not make sense, since it
might
assign a nonzero probability to some $x$ for which\ $f\left(  x,y\right)
$\ is undefined.

\subsection{Applications\label{APPL}}

In this subsection we apply %
\autoref{vardist}\ to prove lower bounds for
two problems of Ambainis. \ To facilitate further research and to
investigate
the scope of our method, we state the problems in a more general way than
Ambainis did. \ Given a group $G$, the \textit{coset problem}
$\operatorname*{Coset}\left(  G\right)  $\ is defined as follows. \ Alice is
given a left coset $C$\ of a subgroup in $G$, and Bob is given an element
$y\in G$. \ Bob must output $1$ if $y\in C$\ and $0$ otherwise. \ By
restricting the group $G$, we obtain many interesting and natural problems.
\ For example, if $p$ is prime then $\operatorname*{Coset}\left(
\mathbb{Z}_{p}\right)  $\ is just the equality problem, so the protocol of
Rabin and Yao \cite{ry}\ yields $Q_{2}^{1}\left(
\operatorname*{Coset}\left(
\mathbb{Z}_{p}\right)  \right)  =\Theta\left(  \log\log p\right)  $.

\begin{theorem}
\label{coset}$Q_{2}^{1}\left(  \operatorname*{Coset}\left(  \mathbb{Z}_{p}%
^{2}\right)  \right)  =\Theta\left(  \log p\right)  $.
\end{theorem}

\begin{proof}
The upper bound is obvious. \ For the lower bound, it suffices to consider a
function $f_{p}$\ defined as follows. \ Alice is given $\left\langle
x,y\right\rangle \in\mathbb{F}_{p}^{2}$\ and Bob is given $\left\langle
a,b\right\rangle \in\mathbb{F}_{p}^{2}$; then%
\[
f_{p}\left(  x,y,a,b\right)  =\left\{
\begin{array}
[c]{ll}%
1 & \text{if }y\equiv ax+b\left(  \operatorname{mod}p\right) \\
0 & \text{otherwise.}%
\end{array}
\right.
\]
Let $\mathcal{B}$ be the uniform distribution over $\left\langle
a,b\right\rangle \in\mathbb{F}_{p}^{2}$, and let $\mathcal{A}_{a,b}$\ be the
uniform distribution over $\left\langle x,y\right\rangle $\ such that
$y\equiv
ax+b\left(  \operatorname{mod}p\right)  $. \ Thus $\mathcal{D}_{1}$\ is the
uniform distribution over $\left\langle x,y\right\rangle \in\mathbb{F}_{p}%
^{2}$; note that%
\[
\Pr_{\left\langle x,y\right\rangle \in\mathcal{D}_{1},\left\langle
a,b\right\rangle \in\mathcal{B}}\left[  f_{p}\left(  x,y,a,b\right)
=0\right]  =1-\frac{1}{p}.
\]
But what about the distribution $\mathcal{D}_{2}$, which is formed by first
drawing $\left\langle a,b\right\rangle \in\mathcal{B}$, and then drawing
$\left\langle x,y\right\rangle $ and $\left\langle z,w\right\rangle
$\ independently from $\mathcal{A}_{a,b}$? \ Given a pair $\left\langle
x,y\right\rangle ,\left\langle z,w\right\rangle \in\mathbb{F}_{p}^{2}$,
there
are three cases regarding the probability of its being drawn from
$\mathcal{D}_{2}$:

\begin{enumerate}
\item[(1)] $\left\langle x,y\right\rangle =\left\langle z,w\right\rangle $
($p^{2}$ pairs). \ In this case%
\begin{align*}
\Pr_{\mathcal{D}_{2}}\left[  \left\langle x,y\right\rangle ,\left\langle
z,w\right\rangle \right]   &  =\sum_{\left\langle a,b\right\rangle
\in\mathbb{F}_{p}^{2}}\Pr\left[  \left\langle a,b\right\rangle \right]
\Pr\left[  \left\langle x,y\right\rangle ,\left\langle z,w\right\rangle
~|~\left\langle a,b\right\rangle \right] \\
&  =p\left(  \frac{1}{p^{2}}\cdot\frac{1}{p^{2}}\right)  =\frac{1}{p^{3}}.
\end{align*}

\item[(2)] $x\neq z$ ($p^{4}-p^{3}$ pairs). \ In this case there exists a
unique $\left\langle a^{\ast},b^{\ast}\right\rangle $\ such that $y\equiv
a^{\ast}x+b^{\ast}\left(  \operatorname{mod}p\right)  $\ and $w\equiv
a^{\ast
}z+b^{\ast}\left(  \operatorname{mod}p\right)  $, so%
\begin{align*}
\Pr_{\mathcal{D}_{2}}\left[  \left\langle x,y\right\rangle ,\left\langle
z,w\right\rangle \right]   &  =\Pr\left[  \left\langle a^{\ast},b^{\ast
}\right\rangle \right]  \Pr\left[  \left\langle x,y\right\rangle
,\left\langle
z,w\right\rangle ~|~\left\langle a^{\ast},b^{\ast}\right\rangle \right] \\
&  =\frac{1}{p^{2}}\cdot\frac{1}{p^{2}}=\frac{1}{p^{4}}.
\end{align*}

\item[(3)] $x=z$ but $y\neq w$ ($p^{3}-p^{2}$ pairs). \ In this case
$\Pr_{\mathcal{D}_{2}}\left[  \left\langle x,y\right\rangle ,\left\langle
z,w\right\rangle \right]  =0$.
\end{enumerate}

Putting it all together,%
\begin{align*}
\left\Vert \mathcal{D}_{2}-\mathcal{D}_{1}^{2}\right\Vert  &  =\frac{1}%
{2}\left(  p^{2}\left\vert \frac{1}{p^{3}}-\frac{1}{p^{4}}\right\vert
+\left(
p^{4}-p^{3}\right)  \left\vert \frac{1}{p^{4}}-\frac{1}{p^{4}}\right\vert
+\left(  p^{3}-p^{2}\right)  \left\vert 0-\frac{1}{p^{4}}\right\vert \right)
\\
&  =\frac{1}{p}-\frac{1}{p^{2}}.
\end{align*}
So taking $\delta=1/p-1/p^{2}$, we have $Q_{2}^{1}\left(
\operatorname*{Coset}\left(  \mathbb{Z}_{p}^{2}\right)  \right)
=\Omega\left(  \log\left(  1/\delta\right)  \right)  =\Omega\left(  \log
p\right)  $ by %
\autoref{vardist}.
\end{proof}

We now consider Ambainis' second problem. \ Given a group $G$ and
nonempty set
$S\subset G$\ with $\left\vert S\right\vert \leq\left\vert G\right\vert /2$,
the \textit{subset problem} $\operatorname*{Subset}\left(  G,S\right)  $\ is
defined as follows. \ Alice is given $x\in G$\ and Bob is given $y\in
G$;\ then Bob must output $1$ if $xy\in S$\ and $0$ otherwise.

Let $\mathcal{M}$\ be the distribution over $st^{-1}\in G$\ formed by
drawing
$s$ and $t$ uniformly and independently from $S$. \ Then let $\Delta
=\left\Vert \mathcal{M}-\mathcal{D}_{1}\right\Vert $, where
$\mathcal{D}_{1}%
$\ is the uniform distribution over $G$.

\begin{proposition}
\label{subset}For all $G,S$ such that $\left\vert S\right\vert
\leq\left\vert
G\right\vert /2$,%
\[
Q_{2}^{1}\left(  \operatorname*{Subset}\left(  G,S\right)  \right)
=\Omega\left(  \log1/\Delta\right)  .
\]

\end{proposition}

\begin{proof}
Let $\mathcal{B}$ be the uniform distribution over $y\in G$, and let
$\mathcal{A}_{y}$\ be the uniform distribution over $x$\ such that
$xy\in S$.
\ Thus $\mathcal{D}_{1}$\ is the uniform distribution over $x\in G$;
note that%
\[
\Pr_{x\in\mathcal{D}_{1},y\in\mathcal{B}}\left[  xy\notin S\right]
=1-\frac{\left\vert S\right\vert }{\left\vert G\right\vert }\geq\frac{1}{2}.
\]
We have%
\begin{align*}
\left\Vert \mathcal{D}_{2}-\mathcal{D}_{1}^{2}\right\Vert  &  =\frac{1}{2}%
\sum_{x,z\in G}\left\vert \frac{\left\vert \left\{  y\in G,s,t\in
S:xy=s,zy=t\right\}  \right\vert }{\left\vert G\right\vert \left\vert
S\right\vert ^{2}}-\frac{1}{\left\vert G\right\vert ^{2}}\right\vert \\
&  =\frac{1}{2}\sum_{x,z\in G}\left\vert \frac{\left\vert \left\{  s,t\in
S:xz^{-1}=st^{-1}\right\}  \right\vert }{\left\vert S\right\vert ^{2}}%
-\frac{1}{\left\vert G\right\vert ^{2}}\right\vert \\
&  =\frac{1}{2}\sum_{x\in G}\left\vert \frac{\left\vert \left\{  s,t\in
S:x=st^{-1}\right\}  \right\vert }{\left\vert S\right\vert ^{2}}-\frac
{1}{\left\vert G\right\vert }\right\vert \\
&  =\frac{1}{2}\sum_{x\in G}\left\vert \Pr_{\mathcal{M}}\left[  x\right]
-\frac{1}{\left\vert G\right\vert }\right\vert \\
&  =\left\Vert \mathcal{M}-\mathcal{D}_{1}\right\Vert \\
&  =\Delta.
\end{align*}
Therefore $\log\left(  1/\delta\right)  =\Omega\left(
\log1/\Delta\right)  $.
\end{proof}

Having lower-bounded $Q_{2}^{1}\left(  \operatorname*{Subset}\left(
G,S\right)  \right)  $ in terms of $1/\Delta$,\ it remains only to
upper-bound
the variation distance $\Delta$. \ The following proposition implies
that for
all constants $\varepsilon>0$, if $S$ is chosen uniformly at random
subject to
$\left\vert S\right\vert =\left\vert G\right\vert ^{1/2+\varepsilon}$, then
$Q_{2}^{1}\left(  \operatorname*{Subset}\left(  G,S\right)  \right)
=\Omega\left(  \log\left(  \left\vert G\right\vert \right)  \right)  $\ with
constant probability\ over $S$.

\begin{theorem}
\label{randset}For all groups $G$ and integers $K\in\left\{  1,\ldots
,\left\vert G\right\vert \right\}  $, if $S\subset G$ is chosen uniformly at
random subject to $\left\vert S\right\vert =K$, then $\Delta=O\left(
\sqrt{\left\vert G\right\vert }/K\right)  $ with $\Omega\left(  1\right)
$\ probability over $S$.
\end{theorem}

\begin{proof}
We have%
\[
\Delta=\frac{1}{2}\sum_{x\in G}\left\vert \Pr_{\mathcal{M}}\left[  x\right]
-\frac{1}{\left\vert G\right\vert }\right\vert \leq\frac{\sqrt{\left\vert
G\right\vert }}{2}\sqrt{\sum_{x\in G}\left(  \Pr_{\mathcal{M}}\left[
x\right]  -\frac{1}{\left\vert G\right\vert }\right)  ^{2}}%
\]
by the Cauchy-Schwarz inequality. \ We claim that%
\[
\operatorname*{EX}_{S}\left[  \sum_{x\in G}\left(  \Pr_{\mathcal{M}}\left[
x\right]  -\frac{1}{\left\vert G\right\vert }\right)  ^{2}\right]  \leq
\frac{c}{K^{2}}%
\]
for some constant $c$. \ {}From this it follows by Markov's inequality that%
\[
\Pr_{S}\left[  \sum_{x\in G}\left(  \Pr_{\mathcal{M}}\left[  x\right]
-\frac{1}{\left\vert G\right\vert }\right)  ^{2}\geq\frac{2c}{K^{2}}\right]
\leq\frac{1}{2}%
\]
and hence%
\[
\Delta\leq\frac{\sqrt{\left\vert G\right\vert }}{2}\sqrt{\frac{2c}{K^{2}}%
}=O\left(  \frac{\sqrt{\left\vert G\right\vert }}{K}\right)
\]
with probability at least $1/2$.

Let us now prove the claim. \ We have%
\[
\Pr_{\mathcal{M}}\left[  x\right]  =\Pr_{i,j}\left[
s_{i}s_{j}^{-1}=x\right]
=\Pr_{i,j}\left[  s_{i}=xs_{j}\right]  ,
\]
where $S=\left\{  s_{1},\ldots,s_{K}\right\}  $\ and $i,j$\ are drawn
uniformly and independently from $\left\{  1,\ldots,K\right\}  $. \ So by
linearity of expectation,%
\begin{align*}
\operatorname*{EX}_{S}\left[  \sum_{x\in G}\left(  \Pr_{\mathcal{M}}\left[
x\right]  -\frac{1}{\left\vert G\right\vert }\right)  ^{2}\right]   &
=\operatorname*{EX}_{S}\left[  \sum_{x\in G}\left(  \left(  \Pr_{i,j}\left[
s_{i}=xs_{j}\right]  \right)  ^{2}-\frac{2}{\left\vert G\right\vert }\Pr
_{i,j}\left[  s_{i}=xs_{j}\right]  +\frac{1}{\left\vert G\right\vert ^{2}%
}\right)  \right] \\
&  =\sum_{x\in G}\left(
\frac{1}{K^{4}}\sum_{i,j,k,l=1}^{K}p_{x,ijkl}\right)
-\frac{2}{\left\vert G\right\vert }\sum_{x\in G}\left(  \frac{1}{K^{2}}%
\sum_{i,j=1}^{K}p_{x,ij}\right)  +\frac{1}{\left\vert G\right\vert }%
\end{align*}
where%
\begin{align*}
p_{x,ij}  &  =\Pr_{S}\left[  s_{i}=xs_{j}\right]  ,\\
p_{x,ijkl}  &  =\Pr_{S}\left[  s_{i}=xs_{j}\wedge s_{k}=xs_{l}\right]  .
\end{align*}

First we analyze $p_{x,ij}$. \ Let $\operatorname*{ord}\left(  x\right)
$\ be
the order of $x$ in $G$. \ Of the $K^{2}$\ possible ordered\ pairs $\left(
i,j\right)  $, there are $K$ pairs with the \textquotedblleft
pattern\textquotedblright\ $ii$\ (meaning that $i=j$), and $K\left(
K-1\right)  $ pairs with the pattern $ij$\ (meaning that $i\neq j$). \ If
$\operatorname*{ord}\left(  x\right)  =1$ (that is, $x$\ is the identity),
then we have $p_{x,ij}=\Pr_{S}\left[  s_{i}=s_{j}\right]  $, so $p_{x,ij}%
=1$\ under the pattern $ii$, and $p_{x,ij}=0$ under the pattern $ij$.\ \ On
the other hand, if $\operatorname*{ord}\left(  x\right)  >1$, then
$p_{x,ij}=0$\ under the pattern $ii$, and $p_{x,ij}=\frac{1}{\left\vert
G\right\vert -1}$\ under the pattern $ij$. \ So%
\[
\frac{1}{K^{2}}\sum_{x\in G}\sum_{i,j=1}^{K}p_{x,ij}=\frac{1}{K^{2}}\left(
K+\left(  \left\vert G\right\vert -1\right)  \frac{K\left(  K-1\right)
}{\left\vert G\right\vert -1}\right)  =1.
\]

Though unnecessarily cumbersome, the above analysis was a warmup for the
more
complicated case of $p_{x,ijkl}$. \ The following table lists the
expressions
for $p_{x,ijkl}$, given $\operatorname*{ord}\left(  x\right)  $\ and the
pattern of $\left(  i,j,k,l\right)  $.%
\[%
\begin{tabular}
[c]{|ll|lll|}\hline
Pattern & Number of such $4$-tuples & $\operatorname*{ord}\left(  x\right)
=1$ & $\operatorname*{ord}\left(  x\right)  =2$ & $\operatorname*{ord}\left(
x\right)  >2$\\\hline
$iiii,iikk$ & $K^{2}$ & \multicolumn{1}{|c}{$1$} & \multicolumn{1}{c}{$0$} &
\multicolumn{1}{c|}{$0$}\\
$ijij$ & $K\left(  K-1\right)  $ & \multicolumn{1}{|c}{$0$} &
\multicolumn{1}{c}{$\frac{1}{\left\vert G\right\vert -1}$} &
\multicolumn{1}{c|}{$\frac{1}{\left\vert G\right\vert -1}$}\\
$ijji$ & $K\left(  K-1\right)  $ & \multicolumn{1}{|c}{$0$} &
\multicolumn{1}{c}{$\frac{1}{\left\vert G\right\vert -1}$} &
\multicolumn{1}{c|}{$0$}\\
$iiil,iiki,ijii,ijjj$ & $4K\left(  K-1\right)  $ &
\multicolumn{1}{|c}{$0$} &
\multicolumn{1}{c}{$0$} & \multicolumn{1}{c|}{$0$}\\
$ijki,ijjk$ & $2K\left(  K-1\right)  \left(  K-2\right)  $ &
\multicolumn{1}{|c}{$0$} & \multicolumn{1}{c}{$0$} &
\multicolumn{1}{c|}{$\frac{1}{\left(  \left\vert G\right\vert -1\right)
\left(  \left\vert G\right\vert -2\right)  }$}\\
$iikl,ijkk,ijik,ijkj$ & $4K\left(  K-1\right)  \left(  K-2\right)  $ &
\multicolumn{1}{|c}{$0$} & \multicolumn{1}{c}{$0$} &
\multicolumn{1}{c|}{$0$%
}\\
$ijkl$ & $K\left(  K-1\right)  \left(  K-2\right)  \left(  K-3\right)  $ &
\multicolumn{1}{|c}{$0$} & \multicolumn{1}{c}{$\frac{1}{\left(  \left\vert
G\right\vert -1\right)  \left(  \left\vert G\right\vert -3\right)  }$} &
\multicolumn{1}{c|}{$\frac{1}{\left(  \left\vert G\right\vert -1\right)
\left(  \left\vert G\right\vert -3\right)  }$}\\\hline
\end{tabular}
\ \
\]
Let $r$\ be the number of $x\in G$\ such that $\operatorname*{ord}\left(
x\right)  =2$, and let\ $r^{\prime}=\left\vert G\right\vert -r-1$\ be the
number such that $\operatorname*{ord}\left(  x\right)  >2$. \ Then%
\begin{align*}
\frac{1}{K^{4}}\sum_{x\in G}\sum_{i,j,k,l=1}^{K}p_{x,ijkl}  &  =\frac{1}%
{K^{4}}\left(
\begin{array}
[c]{c}%
K^{2}+\left(  2r+r^{\prime}\right)  \frac{K\left(  K-1\right)  }{\left\vert
G\right\vert -1}+2r^{\prime}\frac{K\left(  K-1\right)  \left(  K-2\right)
}{\left(  \left\vert G\right\vert -1\right)  \left(  \left\vert G\right\vert
-2\right)  }\\
+\left(  r+r^{\prime}\right)  \frac{K\left(  K-1\right)  \left(  K-2\right)
\left(  K-3\right)  }{\left(  \left\vert G\right\vert -1\right)  \left(
\left\vert G\right\vert -3\right)  }%
\end{array}
\right) \\
&  \leq\frac{1}{\left\vert G\right\vert -3}+O\left(  \frac{1}{K^{2}}\right)
\end{align*}
using the fact that $K\leq\left\vert G\right\vert $.

Putting it all together,%
\[
\operatorname*{EX}_{S}\left[  \sum_{x\in G}\left(  \Pr_{\mathcal{M}}\left[
x\right]  -\frac{1}{\left\vert G\right\vert }\right)  ^{2}\right]  \leq
\frac{1}{\left\vert G\right\vert -3}+O\left(  \frac{1}{K^{2}}\right)
-\frac{2}{\left\vert G\right\vert }+\frac{1}{\left\vert G\right\vert
}=O\left(  \frac{1}{K^{2}}\right)
\]
and we are done.
\end{proof}

{}From fingerprinting we also have the following upper bound. \ Let $q$
be the
periodicity of $S$, defined as the number of distinct sets $gS=\left\{
gs:s\in S\right\}  $ where $g\in G$.

\begin{proposition}
\label{fingerprint}$R_{2}^{1}\left(  \operatorname*{Subset}\left(
G,S\right)
\right)  =O\left(  \log\left\vert S\right\vert +\log\log q\right)  $.
\end{proposition}

\begin{proof}
\begin{sloppypar}
Assume for simplicity that $q=\left\vert G\right\vert $; otherwise we could
reduce to a subgroup $H\leq G$\ with $\left\vert H\right\vert =q$. \ The
protocol is as follows: Alice draws a uniform random prime $p$ from the
range
$\bigl[  \left\vert S\right\vert ^{2}\log^{2}\left\vert G\right\vert
,2\left\vert S\right\vert ^{2}\log^{2}\left\vert G\right\vert \bigr]
$; she
then sends Bob the pair\ $\left(  p,x\operatorname{mod}p\right)  $\
where $x$
is interpreted as an integer. \ This takes $O\left(  \log\left\vert
S\right\vert +\log\log\left\vert G\right\vert \right)  $ bits. \ Bob outputs
$1$ if and only if there exists a $z\in G$\ such that $zy\in S$\ and
$x\equiv
z\left(  \operatorname{mod}p\right)  $. \ To see the protocol's correctness,
observe that if\ $x\neq z$, then there at most $\log\left\vert G\right\vert
$\ primes $p$ such that $x-z\equiv0\left(  \operatorname{mod}p\right)  $,
whereas the relevant range contains $\Omega\left(  \frac{\left\vert
S\right\vert ^{2}\log^{2}\left\vert G\right\vert }{\log\left(  \left\vert
S\right\vert \log\left\vert G\right\vert \right)  }\right)  $\ primes.
\ Therefore, if $xy\notin S$, then by the union bound%
\[
\Pr_{p}\left[  \exists z:zy\in S,x\equiv z\left(  \operatorname{mod}p\right)
\right]  =O\left(  \left\vert S\right\vert \log\left\vert G\right\vert
\frac{\log\left(  \left\vert S\right\vert \log\left\vert G\right\vert
\right)
}{\left\vert S\right\vert ^{2}\log^{2}\left\vert G\right\vert }\right)
=o\left(  1\right)  .
\]
\end{sloppypar}
\end{proof}

\section{Open Problems\label{OPEN}}

\begin{itemize}
\item Are $R_{2}^{1}\left(  f\right)  $ and $Q_{2}^{1}\left(  f\right)
$\ polynomially related for every total Boolean function $f$? \ Also, can we
exhibit \textit{any} asymptotic separation between these measures? \ The
best
separation we know of is a factor of $2$: for the equality function we have
$R_{2}^{1}\left(  \operatorname*{EQ}\right)  \geq\left(  1-o\left(  1\right)
\right)  \log_{2}n$, whereas Winter \cite{winter}\ has shown that $Q_{2}%
^{1}\left(  \operatorname*{EQ}\right)  \leq\left(  1/2+o\left(  1\right)
\right)  \log_{2}n$\ using a protocol involving mixed states.\footnote{If we
restrict ourselves to pure states, then $\left(  1-o\left(  1\right)
\right)
\log_{2}n$\ qubits are needed. \ Based on that fact, a previous version of
this paper claimed incorrectly that $Q_{2}^{1}\left(  \operatorname*{EQ}%
\right)  \geq\left(  1-o\left(  1\right)  \right)  \log_{2}n$.} \ This
factor-$2$ savings is tight for equality: a simple counting argument shows
that $Q_{2}^{1}\left(  \operatorname*{EQ}\right)  \geq\left(  1/2-o\left(
1\right)  \right)  \log_{2}n$; and although the usual randomized
protocol for
equality~\cite{ry}\ uses $\left(  2+o\left(  1\right)  \right)$ $ \log_{2}n$\ bits, there exist\ protocols based on error-correcting codes that
use only
$\log_{2}\left(  cn\right)  =\log_{2}n+O\left(  1\right)  $\ bits. \ All of
this holds for any constant error probability $0<\varepsilon<1/2$.

\item \begin{sloppypar}
As a first step toward answering the above questions, can we
lower-bound
$Q_{2}^{1}\left(  \operatorname*{Coset}\left(  G\right)  \right)  $\ for
groups other than $\mathbb{Z}_{p}^{2}$\ (such as $\mathbb{Z}_{2}^{n}$, or
nonabelian groups)? \ Also, can we characterize $Q_{2}^{1}\left(
\operatorname*{Subset}\left(  G,S\right)  \right)  $\ for all sets $S$,
closing the gap between the upper and lower bounds?
\end{sloppypar}

\item Is there an oracle relative to which $\mathsf{BQP/poly}\neq
\mathsf{BQP/qpoly}$?

\item Can we give oracles relative to which $\mathsf{NP}\cap\mathsf{coNP}%
$\ and $\mathsf{SZK}$\ are not contained in $\mathsf{BQP/qpoly}$? \
Bennett et
al. \cite{bbbv}\ gave an oracle relative to which $\mathsf{NP}\cap
\mathsf{coNP}\not \subset \mathsf{BQP}$, while Aaronson \cite{aar:col}\ gave
an oracle relative to which $\mathsf{SZK}\not \subset \mathsf{BQP}$.

\item Even more ambitiously, can we prove a direct product theorem for
quantum
query complexity that applies to any partial or total function (not just
search)?

\item For all $f$ (partial or total), is $R_{2}^{1}\left(  f\right)
=O\left(
\sqrt{n}\right)  $\ whenever $Q_{2}^{1}\left(  f\right)  =O\left(  \log
n\right)  $? \ In other words, is the separation of Bar-Yossef et al.
\cite{bjk} the best possible?

\item Can the result $D^{1}\left(  f\right)  =O\left(  mQ_{2}^{1}\left(
f\right)  \log Q_{2}^{1}\left(  f\right)  \right)  $\ for partial $f$ be
improved to $D^{1}\left(  f\right)  =O\left(  mQ_{2}^{1}\left(  f\right)
\right)  $? \ We do not even know how to rule out $D^{1}\left(  f\right)
=O\left(  m+Q_{2}^{1}\left(  f\right)  \right)  $.

\item In the Simultaneous Messages (SM) model, there is no direct
communication between Alice and Bob;\ instead, Alice and Bob both send
messages to a third party called the \textit{referee}, who then outputs the
function value. \ The complexity measure is the sum of the two message
lengths. \ Let $R_{2}^{||}\left(  f\right)  $\ and $Q_{2}^{||}\left(
f\right)  $ be the randomized and quantum bounded-error SM complexities
of $f$
respectively, and let $R_{2}^{||,\operatorname*{pub}}\left(  f\right)  $\ be
the randomized SM complexity if Alice and Bob share an arbitrarily long
random
string. \ Building on work by Buhrman et al. \cite{bcww}, Yao
\cite{yao:fing}
showed that $Q_{2}^{||}\left(  f\right)  =O\left(  \log n\right)  $\
whenever
$R_{2}^{||,\operatorname*{pub}}\left(  f\right)  =O\left(  1\right)  $. \ He
then asked about the other direction: for some $\varepsilon>0$, does
$R_{2}^{||,\operatorname*{pub}}\left(  f\right)  =O\left(
n^{1/2-\varepsilon
}\right)  $\ whenever $Q_{2}^{||}\left(  f\right)  =O\left(  \log
n\right)  $,
and does $R_{2}^{||}\left(  f\right)  =O\left(  n^{1-\varepsilon}\right)  $
whenever $Q_{2}^{||}\left(  f\right)  =O\left(  \log n\right)  $? \ In an
earlier version of this paper, we showed that $R_{2}^{||}\left(  f\right)
=O\left(  \sqrt{n}\left(  R_{2}^{||,\operatorname*{pub}}\left(  f\right)
+\log n\right)  \right)  $, which means that a positive answer to Yao's
first
question would imply a positive answer to the second. \ Later we learned
that
Yao independently proved the same result \cite{yao:hw}.

\begin{sloppypar}
Here we ask a related question: can $Q_{2}^{||}\left(  f\right)  $\ ever be
exponentially smaller than $R_{2}^{||,\operatorname*{pub}}\left(
f\right)  $?
\ (Buhrman et al. \cite{bcww}\ showed that $Q_{2}^{||}\left(  f\right)
$\ can
be exponentially smaller than $R_{2}^{||}\left(  f\right)  $.) \ Iordanis
Kerenidis has pointed out to us that, based on the hidden matching
problem of
Bar-Yossef et al. \cite{bjk}\ discussed in %
\secref{PRELIM},\ one can
define a \textit{relation} for which $Q_{2}^{||}\left(  f\right)  $\ is
exponentially smaller than $R_{2}^{||,\operatorname*{pub}}\left(
f\right)  $.
\ However, as in the case of $Q_{2}^{1}\left(  f\right)  $\ versus $R_{2}%
^{1}\left(  f\right)  $, it remains to extend that result to functions.
\end{sloppypar}
\end{itemize}

\section{Acknowledgments}

I thank Oded Regev for pointing out the connection between advice and
one-way
communication; Andris Ambainis for posing the coset and subset problems;
Harry
Buhrman for asking about an upper bound on $\mathsf{BQP/qpoly}$; Lance
Fortnow
for pointing out that $\mathsf{P/poly}\neq\mathsf{PP/poly}$ implies
$\mathsf{PP}\not \subset \mathsf{P/poly}$; Hartmut Klauck and Ronald de Wolf
for discussions that led to improvements of %
\lemref{degreelb}; Iordanis Kerenidis and Andreas Winter for
correcting me about the constant factors in $R_{2}^{1}\left(
\operatorname*{EQ}\right)  $\ and $Q_{2}^{1}\left(
\operatorname*{EQ}\right)  $; and the anonymous reviewers and ToC
editors for helpful comments and corrections.

\bibliographystyle{tocplain}
\bibliography{v001a001}

\pagebreak

\begin{tocauthor}
\begin{tocinfo}[saaronson]
Scott Aaronson\footnote{This work was performed while
the author was a graduate student at UC Berkeley.}\\
postdoc \\
Institute for Advanced Study, Princeton\\
aaronson\tocat{}ias\tocdot{}edu \\
\url{http://www.cs.berkeley.edu/~aaronson}
\end{tocinfo}
\end{tocauthor}
\begin{tocaboutauthor}
\begin{tocabout}[saaronson]
Scott Aaronson received his Ph.D. at the University of
California, Berkeley, in 2004 under
\href{http://www.cs.berkeley.edu/~vazirani}{Umesh Vazirani}.
He is proud to be the
first author in \href{http://theoryofcomputing.org}
{\textsf{Theory of Computing.}}
\end{tocabout}
\end{tocaboutauthor}
\end{document}